\documentclass[superscriptaddress,11pt]{article}
\usepackage{authblk}
\usepackage{array}
\usepackage{type1cm}
\usepackage{lettrine}
\usepackage{graphicx}
\usepackage{geometry}
\usepackage{psfrag}
\usepackage{amsmath}
\usepackage{amsfonts}
\usepackage{appendix}
\usepackage[margin=10pt,font=footnotesize,labelfont=sc,format=hang,labelformat=parens]%
{caption}
\usepackage{subcaption}
\usepackage{float}
\usepackage{accents}
\usepackage{amssymb}%
\providecommand{\U}[1]{\protect\rule{.1in}{.1in}}
\numberwithin{equation}{section}
\def\be{\begin{equation}}
\def\ee{\end{equation}}
\def\ba{\begin{eqnarray}}
\def\ea{\end{eqnarray}}
\def\bi{\begin{itemize}}
\def\ei{\end{itemize}}

\def\D{\mathcal{D}}
\def\e{\epsilon}
\def\k{\kappa}

\begin{document}

\title{Euclidean LQG Dynamics: An Electric Shift in Perspective}

\author{Madhavan Varadarajan}
\affil{Raman Research Institute\\Bangalore-560 080, India}
\maketitle

\begin{abstract}
Loop Quantum Gravity (LQG) is a non-perturbative attempt at quantization of a classical phase space description of
gravity in terms of  $SU(2)$ connections and electric fields. 
As emphasized recently \cite{aame}, 
on this phase space, classical gravitational evolution in {\em time} can
be understood in terms of certain gauge covariant generalizations of Lie derivatives
with respect to a {\em spatial} $SU(2)$ Lie algebra valued vector field called the Electric Shift.
%
We present a derivation of a  quantum dynamics for Euclidean LQG which is informed by this understanding.
In addition to the physically
motivated nature of the action of the Euclidean Hamiltonian constraint so derived, the derivation implies that the spin labels of regulating holonomies
are determined by corresponding labels of the spin network state being acted upon thus eliminating 
the `spin $j$-ambiguity' pointed out by Perez. By virtue of Thiemann's seminal work, the Euclidean quantum dynamics  plays a crucial role
in the construction of the Lorentzian quantum dynamics so that our considerations also have application to Lorentzian LQG.

\end{abstract}

\thispagestyle{empty}
\let\oldthefootnote\thefootnote\renewcommand{\thefootnote}{\fnsymbol{footnote}}
\footnotetext{Email: madhavan@rri.res.in}
\let\thefootnote\oldthefootnote

\section{Introduction \label{sec1}}

Canonical Loop Quantum Gravity \cite{lqgbooks} is an attempt at constructing a non-perturbative canonical quantization of a classical Hamiltonian 
description of gravity in terms of an
%
$SU(2)$ Electric field  and its conjugate connection.
While the electric field bears the interpretation of a spatial 
triad field and hence can be used to construct the spatial geometry, the connection contains information of the extrinsic curvature of the slice as embedded in the  dynamically 
emergent spacetime geometry. The quantum kinematics of Loop Quantum Gravity (LQG) is constructed without recourse to any fixed background  geometry and is by now very well understood. 
The basic kinematic operators are holonomies of the connection along spatial edges and electric fluxes through spatial surfaces and their representation
hints at a picture of discrete quantum spatial geometry.
On the other hand, despite significant progress, a satisfactory treatment of the quantum dynamics of LQG remains an open problem. This
dynamics is driven by the quantum Hamiltonian constraint operator.
While early pioneering efforts \cite{looprepn,abhayjurek} culminating in the 
the seminal work of Thiemann \cite{qsd} demonstrate the existence of such an operator, the construction method yields an operator which is far from unique.\footnote{We note that 
given  the complicated dynamical nature of classical general relativity and the absence of any preferred fixed 
background geometry to regulate operators, the {\em existence}
of a well defined gravitational Hamiltonian constraint operator in LQG is already a remarkable achievement.}
This   operator construction method \cite{qsd} first replaces 
the classical Hamiltonian constraint with a classical approximant built out of classical precursors of the basic  operators used in LQG, replaces these classical
precursors by their quantum correspondents thereby defining a regulated quantum  operator, and in a final step defines the Hamiltonian constraint operator as the limit of 
these regulated operators when the regulating parameter is removed. 
In the last decade, these methods have been applied to, and further developed in, the context
of progressively complicated  toy models which capture more and more aspects of gravity \cite{pft,hk,2+1u13,p1,p2,jureku13,p3}. In these models it turns out that the classical dynamics
in {\em time} can be understood in terms of geometrical deformations of field variables in {\em space}. Progress in the quantum theory is then greatly facilitated by
the incorporation of this property of classical evolution into the quantum dynamics.
%

Recent work \cite{aame} shows that classical time evolution in Euclidean gravity can also be understood in terms of {\em spatial} Lie derivatives, suitably generalised, 
with respect to Lie algebra valued vector field called the {\em Electric Shift}
which is obtained by multiplying the  electric field by the lapse function (which smears the Hamiltonian constraint) and scaling the result by an appropriate power of the determinant of the spatial metric.
In this work we present a 
construction  of  a Hamiltonian constraint operator for Euclidean gravity which attempts to incorporate this central feature of the evolution equations. As a result,  the ensuing quantum dynamics can   
be understood in terms of quantum state transformations generated by a corresponding {\em Electric Shift Operator}. 
The central role of the Electric Shift (and its operator correspondent) in classical (and quantum) dynamics was recognized in earlier work on a toy model system  obtained by 
replacing the $SU(2)$ group in canonical  Euclidean gravity equations  by the abelian group $U(1)^3$ both in 2+1 \cite{2+1u13} as well as in 3+1 dimensions \cite{p1,p2,jureku13,p3}. In the 3+1 dimensional 
case, the classical toy model is exactly that derived by Smolin \cite{leeg0} as  a novel weak coupling limit of Euclidean gravity. 
\footnote{For recent work on the classical theory of this model with a view towards the quantum theory, see \cite{ttu13}.}

In this toy model \cite{2+1u13,p1,p2,p3}, while the starting point was a definition of the quantum dynamics mediated by the Electric Shift, the primary focus was on the construction of
a constraint action which yielded a non-trivial anomaly free constraint algebra. As analysed in detail in \cite{hkt}, the anomaly free property can be seen as an implementation of the 
physical property of {\em spacetime covariance}. Hence, 
while our main focus here is the derivation of Hamiltonian constraint actions for Euclidean gravity  
in which the Electric Shift plays a central role, we shall also try to incorporate features which are necessary for a putative demonstration of consistency of these actions with 
a non-trivial anomaly free quantum constraint algebra. In addition we shall also attempt to incorporate features which seem to facilitate a second physical property beyond 
spacetime covariance, namely the property of {\em propagation} (see References \cite{leeprop,pftprop,u13prop} and, particularly, Reference \cite{ttme}).

The layout of the paper is as follows. 
In section \ref{sec2} we briefly review the 
classical Hamltonian formulation of Euclidean gravity in terms of electric fields and connections
and define  the electric shift. Next we review 
the quantum kinematics underlying LQG. In doing so we find
it useful  to define  holonomies in spin $j$ representations for arbitrary spin $j$.
This `quantum' part of  the material in section \ref{sec2}  is standard and its main purpose is the establishment of  notation rather than pedagogy; indeed, in the rest of the paper, we shall assume familiarity with standard LQG structures.
In section \ref{sec3} we  construct the Electric Shift operator and  provide a heuristic derivation of the 
action of the Hamiltonian constraint in which spin network deformations generated by the Electric Shift operator  play a  central role. 
The derivation relies on a key classical identity connected with the gauge covariant Lie derivative first introduced by Jackiw \cite{jackiw}  to the best of our knowledge.
The final constraint action can also be viewed in the more conventional terms of
holonomy approximants to the curvature rather than deformations generated by the Electric Shift operator. When viewed in this way, it turns out that the holonomy approximants are defined with respect to 
representations of $SU(2)$ which are {\em tailored to the spin labels of the edges of the  spin network being acted upon}. This removes the `spin $j$-ambiguity' in the representation choice
of regulating holonomies raised by Perez \cite{perez}. The end result of section \ref{sec3} is an action which is reminiscent of the standard `QSD' action derived by Thiemann \cite{qsd} in that it acts only at vertices
roughly by adding  a single  extraordinary edge for each pair of edges emanating from the  vertex. The edges are added `one at a time' between each such pair and the result {\em summed} over.
The main  difference here is that, first, the extraordinary edge spin labels are tailored to those of the edges emanating from the vertex rather than being fixed at $j=\frac{1}{2}$ and second,  the way they are
added is slightly different. The resulting constraint action shares the property with its QSD counterpart that a second such constraint action does not act on deformations created by the first.

As first noted in \cite{jplm} in order to obtain a {\em non-trivial} constraint algebra on a `habitat' \cite{lm}, it is necessary for a second constraint action to act on deformations created by the 
first. Such actions were obtained in the 3+1 $U(1)^3$ toy model setting in \cite{p1,p2,p3} wherein state deformations created a new  vertex at which the edges formed a certain conical structure on
which
the second action could (generically) act non-trivially.
Hence, in section \ref{sec4} we develop the action of section \ref{sec3} further so as to obtain the exact analog of these conical deformations but now in the $SU(2)$ context of Euclidean gravity.
In the $U(1)^3$ case such deformations were shown to be consistent with anomaly free constraint commutators \cite{p3}. Preliminary calculations in progress indicate the existence of significant technical complications
when one tries to generalise the considerations of \cite{p3} to the $SU(2)$ case.
Hence in section \ref{sec5} we further develop the action of section \ref{sec4} to obtain, roughly speaking, a mix of actions of
the type presented in sections \ref{sec3}  and \ref{sec4}.  The action seems simple enough  that progress on the `anomaly free' front may be within reach. 
It is also likely that an analysis similar to that of 
\cite{ttme} reveals consistency with propagation.
Section \ref{sec6} is devoted to a discussion of our results and plans for future work.

Unless otherwise specified we shall use units in which $G=\hbar=c=1$. We also choose the  Barbero- Immirzi \cite{giorgio,fer} parameter to be unity. For a first  stab at the problem considered in this work, please
see \cite{laddha}.

\section{\label{sec2} Review of Classical theory and its quantum kinematics }
\subsection{Classical Hamiltonian Formulation}

The phase space variables $(A_a^i, E^a_i)$ are an $su(2)$ connection and conjugate densitized electric field 
on the Cauchy slice $\Sigma$ with $a$ denoting  a tangent space index on $\Sigma$ and $i$ an `internal' index valued in $su(2)$
The $su(2)$ structure constant is denoted by  $\epsilon_i{}^{jk}$. Internal indices are raised and lowered by the $su(2)$ Cartan-Killing metric, denoted by the Kronecker delta symbol $\delta_{ij}$.
The  density weight 2 contravariant metric on $\Sigma$ is defined as $qq^{ab}:= E^a_iE^{bi}$, $q$ being the determinant of the corresponding covariant metric $q_{ab}$.
The action of the gauge covariant derivative ${\cal D}_a$ associated with $A_a^i$ on a  Lie algebra valued scalar $\lambda^i$ is:
\be
\D_a \lambda^i = \partial_a \lambda^i + \epsilon^i_{\;\;jk}A_a^j \lambda^k
\ee
where $\partial_a$ is a flat (coordinate) derivative operator with respect to which the connection is specified.  The curvature of ${\cal D}_a$ is then 
$F_{ab}^{i}:=\partial_{a}{A}_{b}^{i}-\partial_{b}{A}_{a}^{i} +\epsilon^{i}_{jk}A_a^jA_b^k$.

The phase space functions:
\begin{align}
G[\Lambda]  &  =\int\mathrm{d}^{3}x~\Lambda^{i}{\cal D}_{a}E_{i}^{a}\\
D[\vec{N}]  &  =\int\mathrm{d}^{3}x~N^{a}\left(  E_{i}^{b}F_{ab}^{i}-A_{a}%
^{i}{\cal D}_{b}E_{i}^{b}\right) \label{defclassd}\\
H[N]  &  =\tfrac{1}{2}\int\mathrm{d}^{3}x~{N}q^{-1/3}\epsilon^{ijk}E_{i}^{a}E_{j}%
^{b}F_{ab}^{k}, \label{defclassh}
\end{align}
are the Gauss law, diffeomorphism, and Hamiltonian constraints of the theory.
The  Hamiltonian constraint (\ref{defclassh}) is chosen to be of density weight $\frac{4}{3}$ with the lapse $N$ carrying a density weight of $-\frac{1}{3}$.
As discussed in \cite{p1,p2,p3} this choice of density weight is made in anticpation of the putative construction of a {\em non-trivial} anomaly free constraint algebra.
However, our considerations in this paper are independent of this choice of density weight  and (as we shall see, apart from an overall factor of regulating parameter) would go through 
for a Hamiltonian contraint of density weight 1 as well.

The Poisson brackets between the constraints are: 
\begin{align}
\{G[\Lambda],G[\Lambda^{\prime}]\}  &  =\{G[\Lambda],H[N]\}=0\\
\{D[\vec{N}],G[\Lambda]\}  &  =G[\pounds _{\vec{N}}\Lambda]\\
\{D[\vec{N}],D[\vec{M}]\}  &  =D[\pounds _{\vec{N}}\vec{M}] \label{classdd}\\
\{D[\vec{N}],H[N]\}  &  =H[\pounds _{\vec{N}}N]\label{classdh}\\
\{H[N],H[M]\}  &  =D[\vec{L}]+G[A\cdot\vec{L}],\qquad
L^{a}:=q^{-2/3}E_{i}^{a}E_{i}^{b}\left(  M\partial_{b}N-N\partial_{b}M\right) \label{classhh}
\end{align}
The last Poisson bracket (between the Hamiltonian constraints) exhibits
structure functions just as in Lorentzian gravity albeit with a different overall sign.

We  define the Electric Shift $N^a_i$ by
\be
N^a_i = NE^a_i q^{-1/3}
\label{defes}
\ee
As can be checked $N^a_i$ transforms like a Lie algebra valued vector field.

We are concerned exclusively with the Hamiltonian constraint in this paper. Segregating an `Electric Shift'
part in brackets, we write (\ref{defclassh}) as:
\be
H(N) = \frac{1}{2}\int \epsilon^{ijk} {(\frac{N E^a_i}{q^{\frac{1}{3}}})} F_{abk}E^b_j.
\label{hamclass}
\ee

\subsection{Holonomies}
Let $e:[0,s]\rightarrow \Sigma$ be an edge embedded in $\Sigma$.
\footnote{We shall work with the semianalytic category of edges, surfaces and Cauchy slices \cite{lost} although this will not be explicitly needed.}
The classical holonomy function  along the edge $e$ is defined as the path ordered exponential of the connection along the edge $e$. The path ordered exponential, $P\exp -\int_e A $,
is a $2j+1\times 2j+1$  matrix in the spin $j$ representation of the group $SU(2)$ with matrix components defined as follows:
\be
(P\exp -\int_e A)^B_{\;C}:=  \delta^{B}{}_C
+ \sum_{m}(-1)^m  \int_{0}^s ds_1\int_{0}^{s_1}..\int_{0}^{s_{m-1}}ds_m (A(s_1)A(s_2)..A(s_m))^B{}_{C}
\label{pexp}
\ee
where the matrix $A(t)$ is defined through:
\be
A(t)^E_{\;F} := {\dot e}^a(t) A_a^i (e(t)) (\tau^{(j)}_{i})^E_{\;F}
\ee
with $A^i_a(e(t))$ denoting the evaluation of the connection $A_a^i$ at the point on the edge $e$ at parameter value $t \in [0,s]$, 
${\dot e}^a(t)$ denoting the edge tangent at that point and $(\tau^{(j)}_{i})^E_{\;F}$ being the ${}^E_{\;F}$ component of the spin $j$
matrix representative of the $i$th generator  of
the Lie algebra of $SU(2)$.  The holonomy (\ref{pexp}) is then a unitary $2j+1\times 2j+1$ matrix with $B, C \in (-j,-j+1,..,j-1, j)$  and we denote it by $h^{(j)}_e (s,0)$.
In what follows we  shall often omit the $(j)$ superscript to avoid notational clutter.

\subsection{Quantum Kinematics}

Standard $SU(2)$ representation theory implies that all holonomies for  $j> \frac{1}{2}$ can be obtained through suitable linear combinations of products of $j=\frac{1}{2}$ holonomies.
Hence in LQG, while the quantum kinematics supports the action of holonomy operators for all $j$,  it is the standard practice to treat operator correspondents of $j=\frac{1}{2}$ holonomies as fundamental. 
These operators act by multiplication on 
an orthonormal basis of spin network states \cite{carlolee,aajurekarea}. For our purposes a spin network state can be thought of as a wave function of the connection
labelled by a closed oriented graph $\gamma$ consisting of a set of oriented edges $e_{\alpha}, \alpha=1,..,m$ each colored with a spin label $j_{\alpha}$ and a set of  $SU(2)$ invariant tensors, $\{C_v\}$, one at every vertex
$v$
which maps the tensor product of  representations along incoming edges at the vertex to the tensor product of representations along outgoing edges at that vertex.
These tensors are called {\em intertwiners} because they intertwine incoming representations with outgoing ones. We denote the entire set of labels by $S$.
The label set $S$ specifies a corresponding wave function of the connection $S(A)$ as follows.

Let $e(t)$ denote the point on the edge $e$ at parameter value $t$.
By convention we use edge parameterizations such that the beginning point of the edge is at $t=0$ and the final point at $t=1$, the orientation of the edge being in the direction of increasing parameter value.
Evaluate the holonomy of the connection $A_a^i$ along an 
oriented edge $e_{\alpha}$ of the graph in the representation $j_{\alpha}$ to obtain the matrix  $h^{(j_\alpha)}_{e_\alpha}(1,0)^{F_\alpha}_{\;\;E_\alpha}$ so that the index $E_\alpha$ is associated with the beginning point of $e_\alpha$ and $F_\alpha$ with its end point.
The intertwiner at vertex $v$ is a tensor $C_v$ with the following index structure. For the upper index of every incoming edge at $v$, $C_v$ has a corresponding lower index, and 
for the lower  index of every outgoing edge at $v$, it has a corresponding upper index. By virtue of this index structure, the upper and lower indices of $C_v$ can be contracted with lower and upper indices
of outgoing and incoming edges at $v$ respectively. It follows that
the upper and lower indices associated with the tensor product of  intertwiners  over all vertices can then be contracted with corresponding lower and upper indices
of the tensor product of all edge holonomies for all edges of the graph. This contraction then yields a function of the connection $A_a^i$ which is the wave function $S(A)$.  This wave function is gauge
invariant by virtue of the group invariance of the intertwiners.  This group invariance property is as follows.
Consider a spin network $S$ with an $N$ valent vertex at the point $v$. Let the edges at $v$ be outgoing at $v$ so that the 
intertwiner $C$ at $v$ has $N$ upper indices, 
\be
C\equiv C^{A_1..A_N}.
\ee
Our notation implicitly assumes that $A_I$ is valued in the spin $j_I$ representation which colors the $I$th outgoing edge $e_I$ at $v$. Let $g_I$ be the  matrix representative of the $SU(2)$ element $g$  in the $j_I$ representation. 
Then the group  invariance property  of the intertwiner is:
\be
C^{A_1..A_N}g_1^{B_1}{}_{A_1}g_2^{B_2}{}_{A_2}...g_N^{B_N}{}_{A_N} = C^{B_1..B_N}
\nonumber
\ee
We shall use this group invariance property repeatedly in sections \ref{sec3}- \ref{sec5}. In order to reduce notational clutter when doing so we abuse notation slightly, suppress the $I$ index on $g$ and denote the matrix representative
of $g$ also by $g$ (the spin representation of this matrix representative will be clear from context) and write the group invariance property as:
\be
C^{A_1..A_N}g^{B_1}{}_{A_1}g^{B_2}{}_{A_2}...g^{B_N}{}_{A_N} = C^{B_1..B_N}
\label{cgginv}
\ee

By choosing a suitable basis of intertwiners at each vertex, one obtains an orthornormal basis with respect to the $SU(2)$ Haar measure
based Hilbert space measure of LQG (see for example \cite{aajurekarea}).

The Electric field operator can be thought of as acting by functional differentiation on this wave function and its action on any $S(A)$ can be readily inferred from its action on 
a spin $j$ edge holonomy which in turn can be evaluated from (\ref{pexp}). A straightforward computation yields:
\be
{\hat E}^a_i(x) h^{\;A}_{e\;B} (1,0) = i 
\int_{0}^1 dt {\dot e}^a(t) \delta^3(x,e(t)) (h_e(1,t)\tau^{(j)}_i h_e (t,0)) ^{\;A}_{\;\;B} 
\label{ehej}
\ee

This expression is {\em not} a finite linear combination of edge holonomies because of the integral  of the 3 dimensional Dirac delta function along the 1 dimensional edge. Integrating the expression
over a 2 dimensional surface (if necessary, with an intermediate regularization of the delta function \cite{aajurekarea}) yields a well defined {\em Electric flux} operator action which maps
edge holonomies to finite  linear combinations of edge holonomies.
Nevertheless, we shall find it useful to employ the expression 
(\ref{ehej}) in our heuristic argumentation of section \ref{sec3} as well as in our derivation of the Electric Shift operator to which we turn next.

\section{\label{sec3}Derivation of an Electric Shift mediated Hamiltonian constraint action}

In section \ref{sec3.1} we construct the electric shift operator. In section \ref{sec3.2} we display a key identity for our considerations in section \ref{sec3.3}.
In section \ref{sec3.3} we derive the action of the Hamiltonian constraint operator. 
Our arguments in sections \ref{sec3.1} and \ref{sec3.3}  follow the established practice in LQG of defining operators through the following steps: 
(i) first by regulating them, 
(ii) next by defining their regulated action on spin network wave functions
which are assumed to be functions of {\em smooth} connections, (iii) then by 
neglecting terms which are next to leading order in the regulating parameter  and (iv) finally  taking away the regulating parameter to zero.
The key step is (iii)  and the ambiguities in the final operator action in (iv) arise because the neglected terms which
are small for any {\em fixed smooth}   connection argument of the wave function, can be  $O(1)$ in the norm of the  Hilbert space  wherein the wave functions reside.
The procedure is well tested and its heuristic justification stems from the fact that on any fixed graph, the evaluation of a wave function on {\em any} element 
of the quantum configuration space is identical to that on some {\em smooth} connection. While in section \ref{sec3.1} we shall be able to implement step (iv)
in a simple way and obtain an operator on the kinematic Hilbert space, in section \ref{sec3.3} we shall refrain from implementing step (iv) because the 
operator approximant does not have a limit on the kinematic Hilbert space. Instead, the `continuum limit' of step (iv) must be sought in a suitable space of off shell states \cite{p3}
and the construction of such a space represents work in progress.
Additionally in section \ref{sec3.3} we shall regulate a part of the quantum shift operator action of section \ref{sec3.2} in a manner which, while not strictly following (i)-(iii), 
is motivated by the smoothness of classical phase space fields. We shall expand on this cryptic statement  in Footnote \ref{fnshiftregjust} of  section \ref{sec3.3} when we encounter the regulation in question.

\subsection{\label{sec3.1}Electric Shift Operator}

Consider a point $v$ and a coordinate patch $\{x\}$ around $v$.
Fix a coordinate ball $B_{\tau}(v)$ of radius $\tau$ centered at $v$, and
restrict attention to small enough $\tau$ in the following manipulations so
that all constructions happen within the domain of $\{x\}$. At the point $v$, 
define a classical approximant $N_{(\tau)}{}^a_i$ to the Electric shift $N^a_i$ which agrees with $N^a_i$ as $\tau\rightarrow 0$:
\begin{equation}
N_{(\tau)}{}^a_i:=N(x(v)) {q}_{\tau}^{-1/3}\frac{1}{\frac{4\pi
\tau^{3}}{3}}\int_{B_{\tau}(v)}\mathrm{d}^{3}x~{E}_{i}^{a}(x) .
\end{equation}
Here ${q}%
_{\tau}^{-1/3}$ denotes an approximant to ${q}^{-1/3}$ at $v$ so that it agrees with ${q}^{-1/3}$ as $\tau\rightarrow 0$ and $N(x(v))$ denotes the evaluation of the 
density weighted lapse at $v$ in the coordinate system  $\{x\}$.

The corresponding operator, ${\hat N}_{\tau}{}^a_i$ is:
\begin{equation}
{\hat N}_{(\tau)}{}^a_i=N(x(v))\frac{1}{\frac{4\pi
\tau^{3}}{3}}(\int_{B_{\tau}(v)}\mathrm{d}^{3}x~\hat{E}_{i}^{a}(x)) \hat{q}_{\tau}^{-1/3}
\end{equation}
where we have ordered  $\hat{q}_{\tau}^{-1/3}$ to the right. We shall assume that $\hat{q}_{\tau}^{-1/3}$ has the following properties in its action on a spin network:\\
\noindent (i)  For small enough $\tau$,  $\tau^{-2} \hat{q}_{\tau}^{-1/3}$ is independent of $\tau$ and has trivial action if  $v$ is a vertex of the spin network
of valence less than 3.\\
\noindent (ii) When  $\tau^{-2} \hat{q}_{\tau}^{-1/3}$ has non-trivial action at $v$, it only changes the intertwiner at $v$ while leaving the edges at $v$ untouched.

We note that any $\hat{q}_{\tau}^{-1/3}$ deriving  either through Thiemann like identities \cite{qsd,ttbook} involving the Volume operator  or through 
the Tikhonov regularization
\footnote{\label{fntycho} The regularization \cite{eugenio} defines the action of the  inverse of a positive operator ${\hat O}$  on a state $\psi$ as  $\lim_{\e \rightarrow 0^+}  ({\hat O} + \epsilon)^{-2} {\hat O} \psi$
where $({\hat O} + \epsilon)^{-1}$ is defined through spectral analysis. The result is an operator which annhilates the zero eigenvalue states of ${\hat O}$.}
involving  spectral analysis of the Volume operator \cite{eugenio} satisfies properties (i) and (ii) above. 
Consider the action of $\tau^{-2}\hat{q}_{\tau}^{-1/3}$  on a spin network state  $S(A)$ with intertwiner $C$ at its $N$ valent vertex $v$.  This action changes the intertwiner to a new intertwiner 
at $v$ which we denote as $C_{\lambda}$, thereby mapping the spin network to the new spin network $S_{\lambda}(A)$ where $S_{\lambda}$ is obtained from $S$ by replacing $C$
at $v$ by $C_{\lambda}$ at $v$.  It follows that the regulated quantum shift action on $S$ for small enough $\tau$ is:
\ba
     {\hat N}_{(\tau)}{}^a_i S(A)& =& N(x(v))\frac{1}{\frac{4\pi
\tau}{3}}(\int_{B_{\tau}(v)}\mathrm{d}^{3}x~\hat{E}_{i}^{a}(x)) S_{\lambda} \nonumber \\
&=& iN(x(v))\frac{1}{\frac{4\pi\tau}{3}}(\int_{B_{\tau}(v)}d^3x
\sum_{I=1}^N\int_{0}^{1} dt_I {\dot e}_I^a(t) \delta^3(x,e_I(t)) (h_{e_I}(1,t_I)\tau^{(j_I)}_i h_{e_I} (t_I,0)) ^{\;A_I}_{\;\;B_I}   \frac{\partial S_\lambda \;\;\;\;\;\;\;\;\;\;\;\;}{\partial h_{e_I}(1,0)^{\;A_I}_{\;\;B_I}} \nonumber
\\
&=& iN(x(v))\frac{1}{\frac{4\pi\tau}{3}}(\sum_{I=1}^N
\int_{0}^{t_{\tau,I}} dt_I {{\dot e}}^a_I  (h_{e_I}(1,t_I)\tau^{(j_I)}_i h_{e_I} (t_I,0)) ^{\;A_I}_{\;\;B_I}   \frac{\partial S_\lambda \;\;\;\;}{\partial h_{e_I}(1,0)^{\;A_I}_{\;\;B_I}}.
\label{ttau}
\ea
In the second line we have used (\ref{ehej}) to evaluate the action of the triad operator on $S_{\lambda}$. In the third line the $\delta$- function has been integrated over $B_{\tau}(v)$ with the consequence
that the $t_I$ integral is over the part of $I$th edge within $B_{\tau}(v)$ 
so that $t_{\tau, I}$ is the parameter value at which  $e_I$ intersects the 
boundary of $B_{\tau}(v)$. Denoting the unit coordinate tangent vector to the edge $e_I$ at $v$ by ${\hat e}^a_I$, it follows that to leading order in $\tau$ we have:
\be
{\hat N}_{(\tau)}{}^a_i 
S(A) = iN(x(v))\frac{1}{\frac{4\pi\tau}{3}}(\sum_{I=1}^N
\tau {\hat { e}}^a_I  (h_{e_I}(1,0)\tau^{(j_I)}_i  ) ^{\;A_I}{}_{B_I}   \frac{ \partial S_\lambda \;\;\;\;}{ \partial h_{e_I}(1,0)^{\;A_I}{}_{B_I}}
\label{qshifttau}
\ee
Note that the combination $(h_{e_I}(1,0)\tau^{(j_I)}_i  ) ^{\;A_I}_{\;\;B_I}   \frac{\partial S_\lambda \;\;\;\;}{\partial h_{e_I}(1,0)^{\;A_I}_{\;\;B_I}}$ is exactly the result of the action
of the left invariant vector field ${\hat X}_{i\;I}$ of the $I$th copy of $SU(2)$ associated with  the edge holonomy $h_{e_I}(1,0)(A)$  on $S(A)$.
Hence, taking the $\tau\rightarrow 0$ limit of right hand side of (\ref{qshifttau}), we define the action of the quantum shift operator at a vertex $v$ of the spin network $S$ to be
\be
{\hat{N}}^{a}_{j}(v) S(A) =
\frac{3i}{4\pi}N(x(v))
\sum_{I=1}^N
{\hat { e}}^a_I  {\hat X}_{j\;I} S_{\lambda}(A)
\label{qshift}
\ee
From the discussion above, the action of the quantum shift is non-trivial only when the `inverse metric determinant operator' $\lim_{\tau\rightarrow 0}\tau^{-2} \hat{q}_{\tau}^{-1/3}$ has a non-trivial action.
Vertices of $S$ where the inverse metric  determinant operator acts non-trivially will be referred to as {\em non-degenerate} vertices (from (ii) above, such vertices  have a minimum valence of 3).

\subsection{\label{sec3.2}An Important  Identity}

The following identity, and subsequent considerations below will be of use in section \ref{sec2}.
For any vector field $V^a$, it is straightforward to check that the following identity holds:
\be
V^aF_{ab}^i=  {\cal L}_V A_b^i -{\cal D}_b (V\cdot A)^i
\label{vf}
\ee
where ${\cal L}_VA_a^i$ is the Lie derivative of $A_a^i$ with respect to $V^a$:
\be
{\cal L}_V A_a^i = V^b\partial_b A_a^i + A_c^i\partial_b V^c
\ee
Next, define the one parameter family of  $SU(2)$ group elements $g(p,t)$, $p\in \Sigma$ as:
\be
g(p,t) = P\exp - \int_{c(p,t)} A
\label{gpt}
\ee
The path ordered exponential is along the curve $c(p,t)$. The curve $c(p,t)$ is the integral curve of the vector field $V^a$
which starts at the point $p\in \Sigma$ and ends at the point $\phi ({\vec V}, t)\circ p$. Here  $\phi ({\vec V},t)$ is the one parameter family of diffeomorphisms
generated by $V^a$ and parameterised by its affine parameter $t$ (so that $V^a = (\frac{d}{dt})^a$).
It is understood that the  path ordered exponential is evaluated in a spin $j$ representation (we have suppressed the spin $j$ label to avoid notational clutter).
Next, we 
define:
\be
A_a^i(p,t)\tau_i =  g^{-1}(p,t)\;\; (\phi({\vec V},t)_* A_a^i(p))\tau_i\;\; g(p,t)  + g^{-1}(p,t) \partial_a g(p,t)
\label{defagphi}
\ee
where $\phi({\vec V},t)_*$ is the {\em pull back} action of the diffeomorphism $\phi({\vec V},t)$, $\tau_i$ is the $i$th generator of the Lie algebra of $SU(2)$ in the spin $j$ representation
and the summation convention over the repeated index $i$ is implicit.

It is then straightforward to check (as asserted by Jackiw \cite{jackiw}
\footnote{As indicated in Reference \cite{jackiw},  the check consists of (a) expanding  $g(p,t),(\phi({\vec V},t)_* A_a^i(p))$  with $t=\e$ to order $\e$ and neglecting higher order terms,  (b) 
 substituting the resulting expressions with higher order terms so neglected into (\ref{defagphi}),  and (c) substituting  $A^i_b (p,\e)\tau_i$ in (\ref{vdotf}) by  the expression obtained 
 in (c). While it is desireable to also carefully bound the contribution of the neglected  terms  to (\ref{vdotf}), we refrain from attempting to do so here due to the 
 heuristic nature of the arguments which employ (\ref{vdotf}) in our analysis of the constraint action in section \ref{sec3.3} (see Remark (1), section \ref{sec3.3}).
}
) that for small enough $\e$:   
\be
V^aF_{ab}^i \tau_i= \frac{A_b^i(p,\e)\tau_i - A_b^i(p)\tau_i}{\e} + O(\e )
\label{vdotf}
\ee

\subsection{\label{sec3.3}Hamiltonian Constraint Operator Action}

The classical constraint (\ref{hamclass}) has three constituents: the right most triad field, the curvature term and the electric shift.
Following earlier work on toy models \cite{p3} we aim for a constraint action constructed  through the  following 3 step schematic:\\ 
\noindent (a) the combination of the electric shift and the curvature  result in a deformation of the connection $\delta A$ similar to (\ref{vdotf}) \\
\noindent (b) the remaining triad field acts as a functional derivative so that we have, schematically, the combination $\delta A \frac{\delta}{\delta A} \Psi (A)$\\
\noindent (c) the combination $\delta A \frac{\delta}{\delta A} \Psi (A)$ can be approximated in terms of the difference of wave functions:\\
$\Psi (A+\delta A) -\Psi (A)$.

With this `guiding heuristic' in mind, we proceed as follows. We order the quantum shift term to the right and obtain the following regulated constraint action on a spin network $S(A)$:
\ba
{\hat H}(N) S(A) & = &    \frac{1}{2}\int d^3x \epsilon_{i}^{\;\;jk} {\hat F}_{ab}^i \; {\hat E}^b_k \;(N{\hat E}^a_j {\hat q}^{-1/3}) \; S(A)  \nonumber\\
&=& \frac{1}{2}\int d^3x \epsilon_{i}^{\;\;jk} {\hat F}_{ab}^i \; {\hat E}^b_k \;{\hat N}^a_j \; S(A)
\label{3.0}
\ea
where in the  second line we have used the definition of the quantum shift operator.
For simplicity we assume that the  lapse $N$ is of compact support around a  nondegenerate vertex $v$ of $S(A)$, with no other vertex of $S$ in this support.
The action of the right most electric shift operator is then supported only at $v$. In order to obtain a non-trivial action of the constraint operator it is 
necessary to regulate the quantum shift action so as to endow it with a support of non-zero measure. We do this by replacing each edge tangent ${\hat e}^a_I$ in (\ref{qshift})
by a vector field ${\hat e}^a_{I,\epsilon}$ which vanishes outside an open neighbourhood  $U_{\epsilon, I}(v)$ of $v$.
We choose ${\hat e}^a_{I,\epsilon}$, $U_{\epsilon, I}(v)$ such that, for small enough $\e$, the following properties hold.
\\
\noindent (i)    $U_{\epsilon, I}(v)$ is a coordinate ball around $v$ of radius $\e + \e^m$ for some $m>>1$
\\
\noindent (ii) $e_J \cap U_{\epsilon, I}(v) = e_J(t_{\e,J,I}, 0)$ where  $e_J(t_{\e,J,I}, 0)$ is a single segment of $e_J$ running from the vertex $v$ at $t_J=0$ to the point $e_J(t_J= t_{\e,J,I})$.
\\
\noindent (iii)  
Consider the point on $e_I$ which is located at a coordinate distance $\e$ from $v$ along $e_I$. Let the parameter value $t_I$ for this point be $\e_I$.
\footnote{
If $e_I$ was a coordinate straight line, $v$ would be exactly at a distance $\e$ from $e_I(\e_I)$. Since $e_I$ may not be such a straight line, the coordinate distance 
between the points $v, e_I(\e_I)$ is $\e + O(\e^2)$.}
On $e_I$, 
${\hat e}^a_{I,\epsilon}(t_I)  = \lambda(t_I)  {\hat e}^a_I(t_I)$ for some $\lambda (t_I) \geq 0$ with $\lambda (t_I)=1$ for $0\leq t_I \leq \e_I$ .
Thus the regulated edge tangent ${\hat e}^a_{I,\epsilon}(t_I)$ is the same as the unit coordinate edge tangent on $e_I(t_{\e,I,I}, 0)$ 
except for a small neighbourhood of the point $e_I(t_{\e,I,I})$, this neighborhood
being of coordinate size  $O(\e^2)$.
\footnote{If $e_I$ was a coordinate straight line, , from (i) this neighbourhood would be of $O(\e^m)$.}
\\

The action of the regulated quantum shift operator at $v$ is defined to be:
\be
{\hat{N}}^{a}_{\epsilon, j} S(A) =
\frac{3i}{4\pi}N(x(v))
\sum_{I=1}^N
{\hat { e}}^a_{I,\e}  {\hat X}_{j\;I} S_{\lambda}(A)
\label{regqshift}
\ee
This in turn defines the regulated Hamiltonian constraint operator ${\hat H}_{\epsilon}(N)$
\footnote{\label{fnshiftregjust}
Note that this sort of regulation is not of the type (i)-(iii) described in the preamble of section \ref{sec3} as it does not directly involve connection dependent terms.
However the necessity of that regulation as well as this one both arise due to the distributional nature of the quantum fields.  More in detail the regulation (i)-(iii) is 
necessary because the distributional nature of elements of the quantum configuration space of connections only allow holonomies to be defined as opposed to 
the local connection fields of the classical theory. Similarly the  regulation of the quantum shift is necessitated by virtue of the distributional nature of the action of the quantum electric field operator (\ref{ehej})
and that of the inverse determinant of the metric operator (see (i), (ii) of section \ref{sec3.1}) which is in contrast to the smoothness of the corresponding classical fields.
Note that this behavior of these electric field dependent operators is also, ultimately, a consequence of the distributional nature of elements of the quantum configuration space of connections. Finally, we note that 
the regulation has effectively the same consequence as that of conventional regulations of the Hamiltonian constraint which, despite the non-triviality of the action of the inverse metric determinant exclusively
at isolated points, yield contributions to the regulated operator action from small neighbourhoods of these points \cite{qsd}
}:
\ba
{\hat H}_{\epsilon}(N) S(A)& = & \frac{1}{2}\int d^3x \epsilon_{i}^{\;\;jk} {\hat F}_{ab}^i \; {\hat E}^b_k \; {\hat{N}}^{a}_{\epsilon, j} \; S(A)  \nonumber\\
&=& \frac{3i}{8\pi}N(x(v))
\int_{U_{\epsilon,I}(v)} d^3x \epsilon_{i}^{\;\;jk} {\hat F}_{ab}^i \; {\hat E}^b_k \;\sum_{I=1}^N
{\hat { e}}^a_{I,\e}  {\hat X}_{j\;I} S_{\lambda}(A)
\nonumber \\
&=& \frac{3i}{8\pi}N(x(v)) \sum_{I} \int_{U_{\epsilon, I}(v)} d^3x \epsilon_{i}^{\;\;jk} ({\hat e}^a_{I,\epsilon}{\hat F}_{ab}^i)  {\hat E}^b_k  \; {\hat X}_{j,I}S_{\lambda}(A) 
\ea

Next we act with  ${\hat E}^b_k$ through (\ref{ehej}), and integrate the Dirac delta function in that action over $U_{\epsilon, I}(v)$. The result is, similar to 
the integral in (\ref{ttau}), a sum over integrals one for each edge with the $J$th integral being over   the part of $J$th edge within  $U_{\epsilon, I}(v)$.
Recalling that $t_{\e, J,I}$ is  the parameter value at which  $e_J$ intersects the 
boundary of $U_{\e, I}(v)$,  we have:
\ba
&{\hat H}_{\epsilon}(N) S(A)
= \frac{3i}{8\pi}N(x(v)) \sum_{I} \int_{U_{\epsilon, I}(v)} d^3x \epsilon_{i}^{\;\;jk} ({\hat e}^a_{I,\epsilon}{\hat F}_{ab}^i)  {\hat E}^b_k  \; {\hat X}_{j,I}S_{\lambda}(A) \;\;\;\;\;\;\;\;\;\;\;\;\;\;\;\;\;\;\;\;\;\;\;\;\;\;\;\;
\nonumber \\
&= -\frac{3}{8\pi}N(x(v))  \sum_{I,J }\int_{0}^{t_{\e,J,I}} dt_J  ({\hat e}^a_{I,\epsilon}{\hat F}_{ab}^i) {\dot e}^b_{J}(t_J)
                                      \e_i^{\;\;jk}[ h_{e_J}(1,t_J) \tau_k h_{e_J}(t_J,0)]^{A_J}_{\;\;B_J} \frac{\partial{\hat X}_{j,I}S_{\lambda}(A) }{\partial h_{e_J\;B_J}^{A_J}} \nonumber \\
&= -\frac{3}{8\pi}N(x(v))  \sum_{I,J }\int_{0}^{t_{\e,J,I}} dt_J  ({\hat e}^a_{I,\epsilon}{\hat F}_{ab}^i) {\dot e}^b_{J}(t_J)
                                      \e_i^{\;\;jk} h_{e_J}(1,0) ({\tau_k} + O(\e ))^{A_J}_{\;\;B_J}  \frac{\partial{\hat X}_{j,I}S_{\lambda}(A) }{\partial h_{e_J\;B_J}^{A_J}} \nonumber\\
& = -\frac{3}{8\pi} N(x(v))  \sum_{I}\sum_{J\neq I }\int_{0}^{t_{\e,J,I}} dt_J  ({\hat e}^a_{I,\epsilon}{ F}_{ab}^i) {\dot e}^b_{J}(t_J)
                                      \e_i^{\;\;jk} ( h_{e_J}(1,0) \tau_k)^{A_J}_{\;\;B_J}  \frac{\partial{\hat X}_{j,I}S_{\lambda}(A) }{\partial h_{e_J\;B_J}^{A_J}} \;\;\;\;\;\;\;\;\;\;\;\;\;\;\;\;
\label{3.1}
\ea                                      
where in the last line we have only retained the leading order contribution in $\e$ ( see (ii) in the preamble of section \ref{sec3}).
Further, since ${\hat F}_{ab}^i$ is a connection dependent operator, it acts by multiplication. Hence we have removed the `hat' on it in the last line.
Finally, note that if $I=J$ we get no contribution by virtue of the antisymmetry of $F_{ab}^i$ together with the fact (see (iii) above of this section)  that  ${\hat e}^a_{I,\epsilon}$ is parallel to ${\dot e}^a_{I}$ along the $I$th edge.
Hence, in the last line the second sum is over $J\neq I$. Note that $\tau_k$ is in the representation $j_J$ which labels the edge $e_J$.

Next, we replace $\e_i{}^{jk}\tau_k$ by the commutator between $\tau_i$ and $\tau_j$ to create  the combination 
$({\hat e}^b_{I,\epsilon}{ F}_{ab}^i) \tau_i$ so as to facilitate the application of (\ref{vdotf}).
\footnote{\label{fntaurep}
Note that $\tau_i, \tau_j$ in this commutator are matrices in the spin $j_J$ representation which labels $e_J$ because $\tau_k$ is in this representation.
} We obtain:
\be
{\hat H}_{\epsilon}(N) S(A)
= -\frac{3}{8\pi}N(x(v))  \sum_I\sum_{J\neq I }  ( h_{e_J}(1,0)   \int_{0}^{t_{\e,J,I}} dt_J  
  {\dot e}^b_{J} [ ({\hat e}^a_{I,\epsilon}{ F}_{ab}^i\tau_i , \tau^j ])^{A_J}_{\;\;B_J}
\frac{\partial{\hat X}_{j,I}S_{\lambda}(A) }{\partial h_{e_J\;B_J}^{A_J}} 
\label{3.2}
\ee
Applying (\ref{vdotf}) with $V^a={\hat e}^a_{I,\epsilon}$ to the  part of (\ref{3.2}) which depends non-trivially on the variable of integration $t_J$, and suppressing matrix indices, 
we have to leading order in $\e$ that:
\be
\int_{0}^{t_{\e,J,I}} dt_J   {\dot e}^b_{J} {\hat e}^a_{I,\epsilon}{ F}_{ab}^i\tau_i
= \int_{0}^{t_{\e,J,I}} dt_J   {\dot e}^b_{J} \left(\frac{A_b^i(e_J(t_J),\e)\tau_i - A_b^i(e_J(t_J))\tau_i}{\e} + O(\e) \right),
\label{3.3}
\ee
where the point on the edge $e_J$ at parameter value $t_J$ is denoted by $e_J(t_J)$
Since the integral is over a small  initial part of the $J$th edge of coordinate size of $O(\e)$, we replace the difference between the integrated connections
by their holonomies. Accordingly, suppressing matrix indices, it is straightforward to check  that:
\ba
-(\int_{0}^{t_{\e,J,I}} dt_J   {\dot e}^b_{J}A_b^i(t_J,\e)\tau_i -  \int_{0}^{t_{\e,J,I}} dt_J    {\dot e}^b_J A_b^i(t_J)\tau_i)
=
 h_{e_J}(t_{\e,J,I}, 0)(A_{\e})-  h_{e_J}(t_{\e,J,I}, 0)(A)  + O(\e^3) \nonumber \\
=\left[(h_{e_J}(t_{\e,J,I}, 0)   (A))^{-1} \big(h_{e_J}(t_{\e,J,I}, 0)(A_{\e})-  h_{e_J}(t_{\e,J,I}, 0)(A) \big)\right]  + O(\e^3).\;\;\;\;\;
\label{3.4}
\ea
Here $A_{\e}$ denotes the connection given by setting $V^a={\hat e}^a_{I,\epsilon}$  and $t=\epsilon$ in  (\ref{defagphi}).
\be
h_{e_J}(t_{\e,J,I}, 0)(A_{\e}) = (g(e_J(t_{\e,J,I}), \e)^{-1} h_{e_J} (t_{\e,J,I}, 0)    ((\phi_{I,\e})_*(A)) g(v, \e)
\ee
where we have denoted  $\phi({\vec V},\epsilon)|_{V^a={\hat e}^a_{I,\epsilon}}$  in (\ref{defagphi}) by $\phi_{I,\e}$.
Recall from (\ref{gpt}) that $g(p,\e)$ in the above equation is the holonomy along the integral curve of the vector field 
$V^a={\hat e}^a_{I,\epsilon}$  from $p$ to the point $\phi_{I,\e}(p)$. Since this vector field vanishes outside $U_{\e, I}(v)$ and since
$e_J(t_{\e,J,I})$  is the point at which $e_J$ intersects the boundary of $U_{\e,I}(v)$, it follows that $g(e_J(t_{\e,J,I}, \e) = {\bf 1}$. Next, note that by virtue of property (iii) of 
${\hat e}^a_{I,\epsilon}$  above,  the integral curve of ${\hat e}^a_{I,\epsilon}$ starting at $v$ runs along $e_I$. It follows from (iii) that 
\be
g(v, \e) = h_{e_I}(\e_I,0)(A) .
\ee
where the parameter value $t_I=\e_I$ is such that 
the coordinate distance from $v$ to $e_I( \e_I)$ is $\e$ to $O(\e^2)$.
Finally note that 
\be
h_{e_J}(t_{\e,J,I}, 0)(\phi_{I,\e}*(A)) = h_{\phi_{I,\e}(e_J)}(t_{\e,J,I}, 0)(A).
\ee
Here $\phi_{I,\e}(e_J)$ denotes the image of  the edge $e_J$ by $\phi_{I,\e}$ and we have used the natural parameterization of $\phi_{I,\e}(e_J)$ induced by $\phi_{I,\e}$ from $e_J$. Explicitly, in this parameterization the point 
$(\phi_{I,\e}(e_J))(t_J)$ on $\phi_{I,\e}(e_J)$ with parameter $t_J$ is the image $\phi_{I,\e}( e_J(t_J))$ of the point on $e_J$ with parameter $t_J$.

Putting all this together in (\ref{3.4}), we get:
\ba
{}&-&\left(\int_{0}^{t_{\e,J,I}} dt_J   {\dot e}^b_{J}A_b^i(t_J,\e)\tau_i -  \int_{0}^{t_{\e,J,I}} dt_J    {\dot e}^b_J A_b^i(t_J)\tau_i \right)\nonumber \\
&=&\left[(h_{e_J}(t_{\e,J,I}, 0)   )^{-1} \big(  h_{\phi_{I,\e}(e_J)}  (t_{\e,J,I}, 0)     h_{e_I}(\e_I,0)    -  h_{e_J}(t_{\e,J,I}, 0) \big)\right]  + O(\e^3) \nonumber\\
&=& (h_{e_J}(t_{\e,J,I}, 0)   )^{-1} h_{\phi_{I,\e}(e_J)} (t_{\e,J,I}, 0)
h_{e_I}(\e_I,0) - {\bf 1}+ O(\e^3) \nonumber\\
&=& h_{l_{IJ,\e}} - {\bf 1}+ O(\e^3) 
\label{3.5}
\ea
where all holonomies are evaluated with respect to the connection $A$ and where we have defined the loop $l_{IJ,\e}$ as:
\be
l_{IJ,\e} = ({e_J}_{(t_{\e,J,I}, 0)}   )^{-1}\circ  \phi_{I,\e} ({e_J}_{(t_{J,I,\e},0)}    ) \circ {e_I}_{ (\e_I,0) } 
\label{deflij}
\ee
where the subscripts indicate the beginning and end values of the parameter values for each of the edges in the equation.

\begin{figure}
\centering
\begin{subfigure}[h]{0.3\textwidth}
    \includegraphics[width=\textwidth]{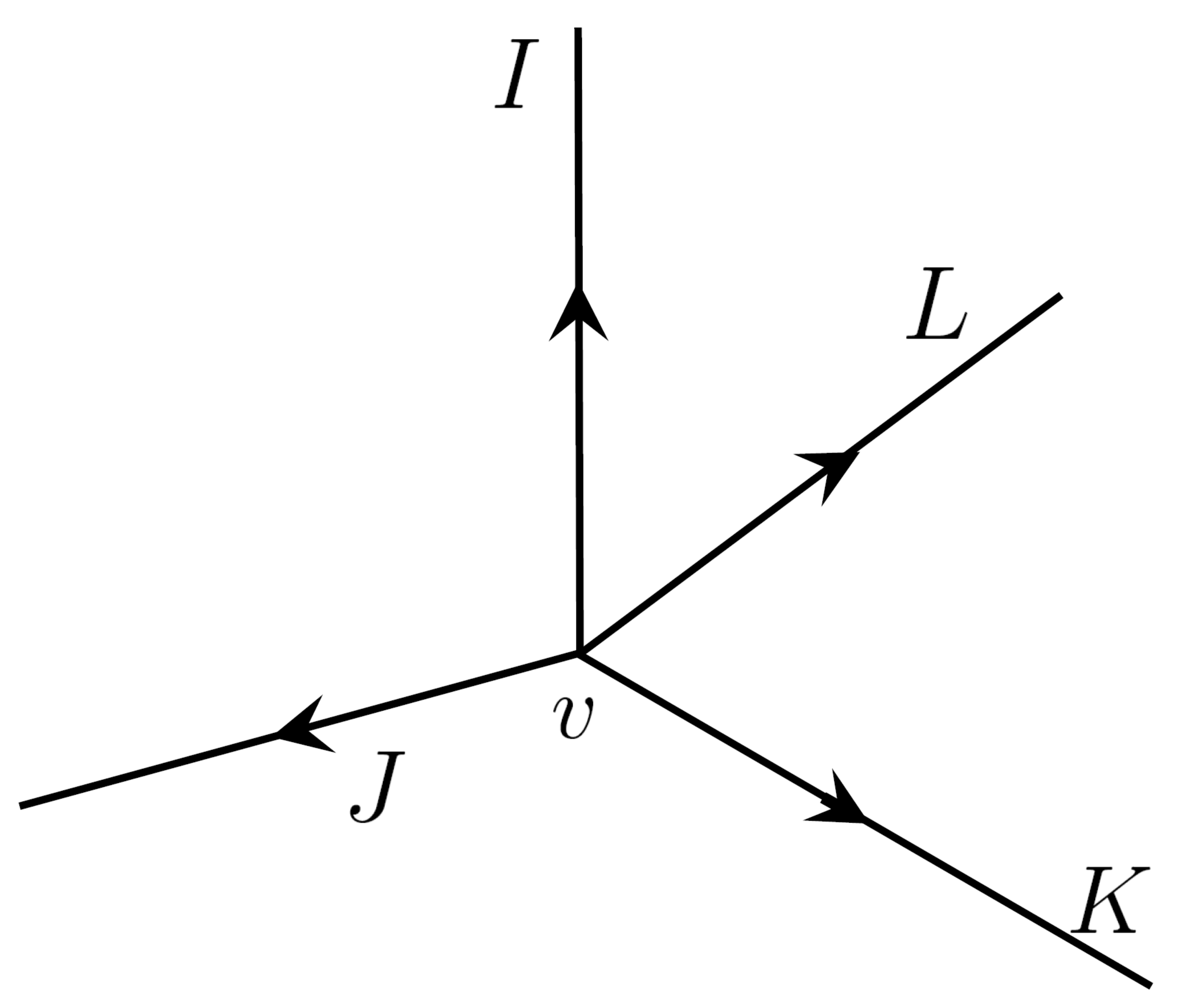}
    \caption{}
 \label{undef}
  \end{subfigure} \quad
  \begin{subfigure}[h]{0.3\textwidth}
    \includegraphics[width=\textwidth]{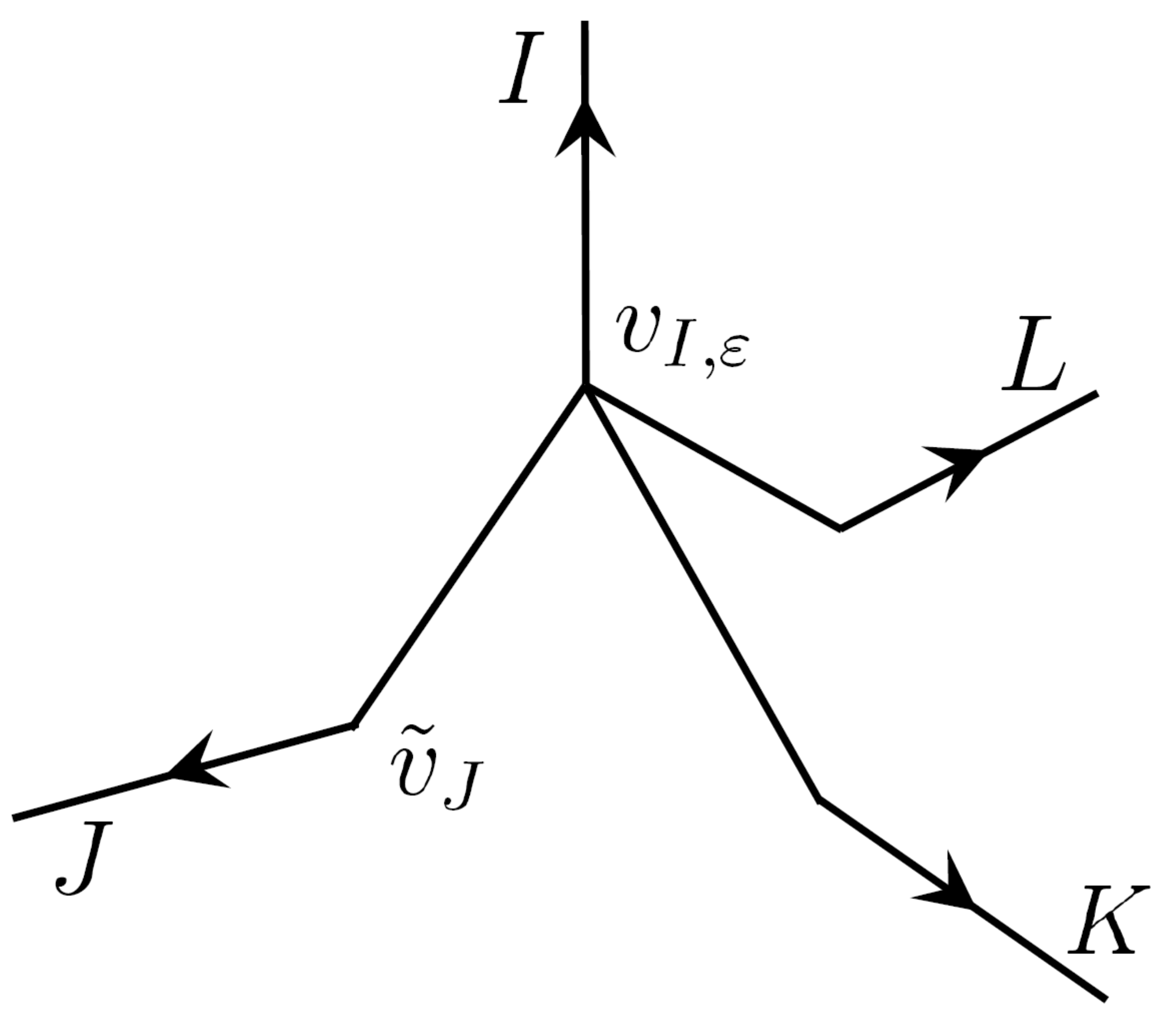}
    \caption{}
   \label{condef}
  \end{subfigure}
\begin{subfigure}[h]{0.10\textwidth}
    \includegraphics[width=\textwidth]{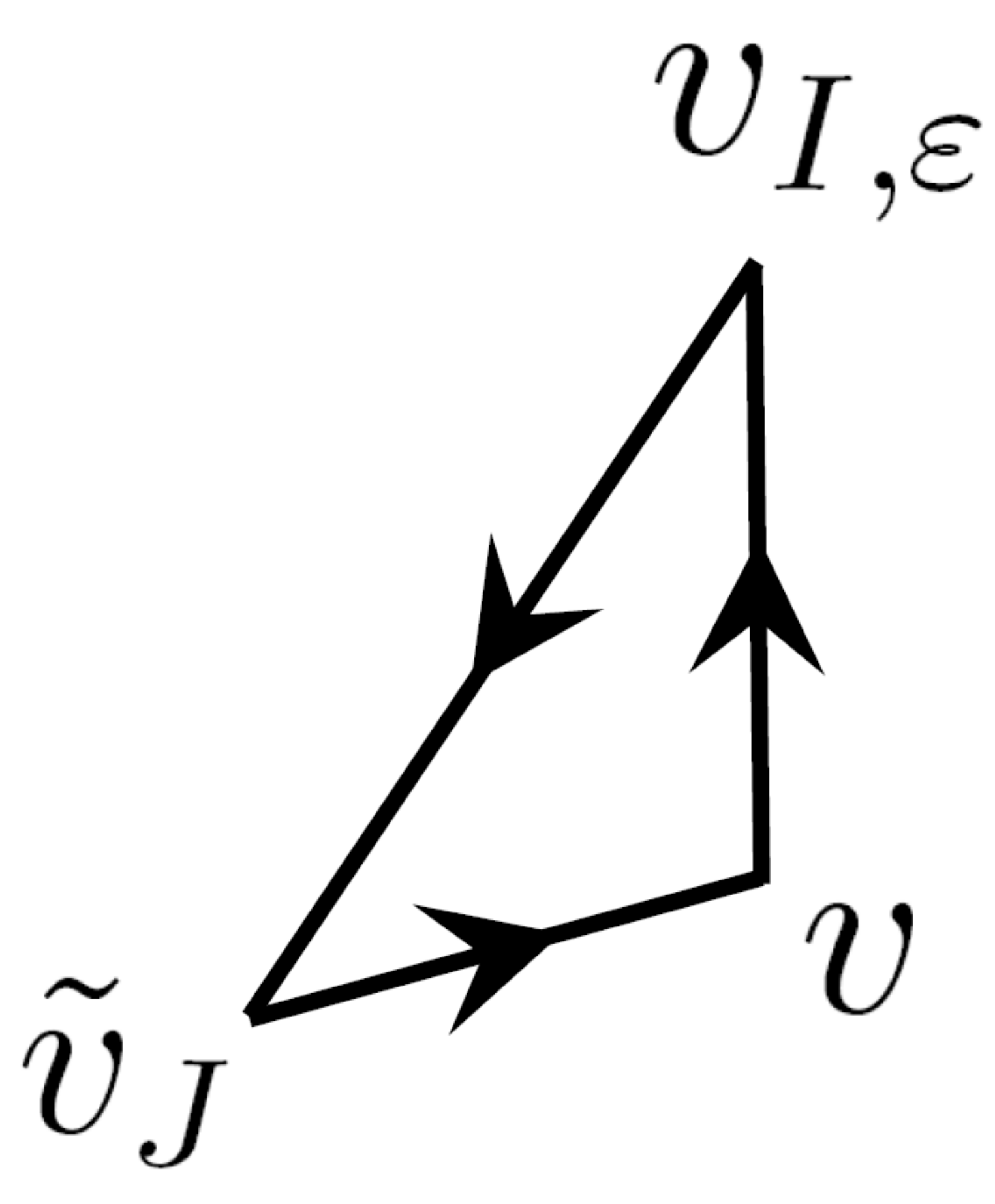}
    \caption{}
   \label{figloop}
  \end{subfigure} \quad
  \caption{ Fig \ref{undef} shows the undeformed vertex structure at $v$. The vertex structure is deformed by $\phi_{I,\e}$ along its $I$th edge in Fig \ref{condef} wherein the displaced
vertex $v_{I,\e}$ and intersection point ${\tilde v}_J$ between the $J$th edge and its deformed image are labelled. Fig \ref{figloop} shows the loop 
$l_{IJ,\e}$ which starts from $v$,  runs along the $I$th edge to $v_{I,\e}$, moves along the $J$th displaced edge to ${\tilde v}_J$ and then back to $v$ along (and in the opposite
direction to) $e_J$.
}%
\label{fig1}%
\end{figure}

As shown in Fig \ref{figloop} this loop starts at $v$, goes along $e_I$ till it reaches the  displaced image $\phi_{I,\e}(v)$ of $v$,  then runs along the displaced $J$th edge
$\phi_{I,\e}({e_J})$ till it hits $e_J$,  and then runs back along $e_J$ in the incoming direction to $v$.
From Footnote \ref{fntaurep} it follows that  $h_{l_{IJ,\e}}$ in (\ref{3.5}) is a holonomy in the spin $j_J$ representation and that ${\bf 1}$ in (\ref{3.5}) is the identity in the spin $j_J$ representation.
Using (\ref{3.5}) and (\ref{3.3}) in (\ref{3.2}) and dropping the next to leading order term deriving from the $O(\e^3 )$ term in (\ref{3.5}), we obtain our final result:
\ba
{\hat H}_{\epsilon}(N) S(A)
&=& \frac{3}{8\pi}N(x(v)) \frac{1}{\e} \sum_{I\neq J } ( h_{e_J}(1,0) [(h_{l_{IJ,\e}} - {\bf 1}), \tau^j ])^{A_J}_{\;\;B_J} 
\frac{\partial{\hat X}_{j,I}S_{\lambda}(A) }{\partial h_{e_J\;B_J}^{A_J}} 
\label{3.6}\\
&=& \frac{3}{8\pi}N(x(v)) \frac{1}{\e} \sum_{I\neq J } ( h_{e_J}(1,0) [h_{l_{IJ,\e}}, \tau^j ] )^{A_J}_{\;\;B_J}. 
\frac{\partial{\hat X}_{j,I}S_{\lambda}(A) }{\partial h_{e_J\;B_J}^{A_J}} 
\label{3.7} 
\ea
We make the following remarks:\\

\noindent (1) Equation  (\ref{vdotf}) involves the approximation of a Lie derivative by a small finite diffeomorphism minus the identity. This is valid for vector fields which are 
 {\em independent} of the smallness parameter $\e$. However, in our considerations the relevant vector fields $\{{\hat e}^a_{I,\e},I=1,..,N\}$ have $\e$ dependent supports so that our application of (\ref{vdotf})
is not, strictly speaking, valid. We feel that this sloppiness is excusable at the level of heuristics employed in our arguments. Indeed, since a similar  logical jump
was made in our reasonably successful treatment of toy models \cite{p1}, we are inclined to treat  the finite deformation maps $\phi_{I,\e}$ as primary with the vector fields $\{{\hat e}^a_{I,\e}\}$
serving as crutches to be discarded. 
\\

\noindent (2) While our argumentation was based on the finite deformations generated by the `vector field' part of the quantum shift, the final result (\ref{3.6}) can be 
naturally interpreted in terms of electric field operator actions and curvature approximants (as is usually done) as follows. Recall our starting point:
\be
{\hat H}(N) S(A)  = \frac{1}{2}\int d^3x \epsilon_{i}^{\;\;jk} {\hat F}_{ab}^i \; {\hat E}^b_k \;(N{\hat E}^a_j {\hat q}^{-1/3}) \; S(A) .
\ee
Upto factors of $\e$ the successive action of each term in this expression can be seen to result in (\ref{3.6}).  
The operator ${\hat q}^{-1/3}$ maps $S$ to $S_{\lambda}$ in (\ref{3.6}). The lapse $N$ is evaluated at $v$ and so appears as an overall factor in (\ref{3.6}).
The ${\hat E}^a_j$  inserts a $\tau_j$ on each edge $e_I$ at $v$ in the spin $j_I$ representation which colors the edge $e_I$,
together with a factor of ${\hat e_I}^a$. The $\tau_j$ insertion corresponds to the ${\hat X}_{j,I}$ operator  in (\ref{3.6}).
The ${\hat E}^b_k$ operator likewise inserts $\tau_k$ in the spin $j_J$ representation on each edge $e_J$ at $v$ (thus generating the derivatives with respect to $h_{e_J}$ in (\ref{3.6})) and generates factors of ${\hat e}^b_J$. 
The $\epsilon_i^{\;\;jk}$  term converts this inserion of  $\tau_k$ into  a commutator of $\tau_i$ and $\tau_j$, also in the spin $j_J$ representation. The $\tau_i$ together with the edge tangents generated by the two electric field operators
combines with the   $F_{ab}^i$ term to give ${\hat e_I}^a {\hat e}^b_J F_{ab}^i\tau_i$. This term can be naturally approximated by the difference of the holonomy of a small loop and the identity (both in the $j_J$ representation)
such that the small loop has a vertex at $v$ with sides along $e_I$ and $e_J$ and a third side opposite $v$. This is exactly the $h_{l_{IJ,\e}} - {\bf 1}$ term in (\ref{3.6}).
Since the $\tau_i$ occurs as a commutator with $\tau_j$ and since these are both insertions on $e_J$, we get exactly the combination  $(h_{e_J}(1,0) [(h_{l_{IJ,\e}} - {\bf 1}), \tau^j ])^{A_J}_{\;\;B_J}$ in (\ref{3.6}).

We can now count the factors of $\e$. The ${\hat q}^{-1/3}$ operator would correspond to $(\frac{ {\hat V} }{\e^3})^{-\frac{2}{3}}$ where ${\hat V}$ is the volume operator for a small region of coordinate size $\e^3$.
This yields a factor of $\e^2$ (times a finite operator whose action we have already accounted for above). Similarly since electric fluxes are well defined operators each electric field operator 
contributes $\e^{-2}$. The curvature  is approximated by a small loop holonomy minus the identity divided by the small loop area so it contributes $\e^{-2}$. Finally, the measure contributes $\e^3$ resulting in 
an overall factor of $(\e)^{2-4-2+3}= \e^{-1}$ which is what we have. The remaining factor of $\frac{3}{4\pi}$ comes from our particular choice of regularization.

We emphasize that despite this interpretation in terms of curvature approximants, our viewpoint and intuition based on the considerations of \cite{aame,p3}, is to treat the finite deformations $\phi_{I,\e}$ as the fundamental building blocks of 
the constraint action.
\\

\noindent (3) Equation (\ref{3.7}) tells us that the Hamiltonian constraint acts at non-degenerate vertices and modifies the vertex structure by the addition of the small `triangular'  loop
$l_{I,J,\e}$. The  `triangle' has a vertex at $v$ from which 2 of its sides emanate along $e_I$ and $e_J$, the third side joining the vertex of the triangle on $e_I$ with that on $e_J$.
This sort of a modification is very similar to that engendered by Thiemann's Quantum Spin Dynamics (QSD) construction of the Hamiltonian constraint \cite{qsd} with the third side here corresponding to an extraordinary edge in \cite{qsd}.
As in the QSD case. the two new vertices are planar and at most trivalent, and from (5) below,  gauge invariant. 
It  follows that with respect to standard constructions of the inverse determinant of the metric, these vertices
are degenerate. Hence, just as in the QSD case   a second constraint would not  see these new vertices. Note also that our treatment is general enough that it goes through for {\em any} density
weight of the constraint (apart from density 2 for which there is no factor involving the determinant of the metric). 
All that changes is the intertwiner $C_{\lambda}$ at $v$ and the overall power of $\e$. In particular for the density one case, as in the QSD case, there is no overall factor of $\e$.
\\

\noindent (4)  Despite these similarities, the constraint action (\ref{3.7}) differs from the QSD constraint in two ways. First, the extraordinary edge (and the loop $l_{I,J,\e}$) in (\ref{3.7}) is 
colored by the spin label of the $J$th edge in contrast to the QSD case in which the color is fixed at $j=\frac{1}{2}$. Since the coloring is fixed by our argumentation and not put in by hand, 
and since the small loop holonomy term serves as a curvature approximant by virtue of our discussion in (2) above, it follows
that our derivation completely fixes the spin representation ambiguity pointed out by Perez \cite{perez}. The second way in which our action differs from that of the QSD constraint is in its implementation 
of the curvature term in terms of holonomy {\em composition}. Here the term combines with matrix representatives of the Lie algebra so as to obtain a result in which holonomies are {\em composed}. In contrast, standard treatments, including that of 
the QSD constraint approximate the curvature term in terms of small loop  holonomy {\em traces} with respect to matrix representatives of the Lie algebra and the result is then a {\em multiplication} by such traces instead of 
the holonomy {\em composition} seen in (\ref{3.7}). In our view, it is this composition which most directly allows the final result (\ref{3.6}) to be interpreted as a precise implementation of our 
guiding heuristic ``$\Psi (A+\delta A) -\Psi (A)$'' (see (c) at the beginning of this section).
\\

\noindent (5) It is straightforward to check that the expression (\ref{3.7}) is gauge invariant with respect to gauge transformations of the connection.
To do so note first that the expression is explicitly gauge invariant with respect to gauge transformations away from $v$ by virtue of the gauge invariance of $S_{\lambda}$ together with the fact that 
all the $\tau$ matrix insertions are only at $v$. Next consider gauge transformations which are non-trivial at $v$.  Let $g\in SU(2)$ be a gauge transformation at $v$. We shall slightly abuse notation and denote its matrix representative in the
spin $j_I$ representation also as $g$. It will be clear from the context as to which spin representation $g$ is valued in. 
Suppressing matrix indices, the holonomy $h_e$ for an incoming edge at $v$ transforms as $h_e \rightarrow g^{-1}h_e$, for a outgoing edge $e$ at $v$ as $h_e\rightarrow h_e g$ and for a  loop $l$  starting in an outgoing direction at $v$
and ending  in an incoming direction at $v$, $h_l\rightarrow g^{-1}h_l g$. 
We write $S_{\lambda}(A)$ as:
\be
S_{\lambda} (A) = C_{\lambda}^{A_1..A_N} (\prod_{I=1}^N h_{e_I}(1,0)^{B_I}{}_{A_I} ) S_{rest}{}_{B_I..B_N}(A)
\label{sbrest}
\ee
where $S_{rest}$ depends on edge holonomies of edges which do not emanate from  $v$.  
As mentioned earlier we have  assumed that all edges at $v$ in $S$ are {\em outgoing}. It is immediate to check that the gauge invariance of $S_{\lambda}(A)$ is then a consequence of the group invariance property of the 
intertwiner $C_{\lambda}$ (see (\ref{cgginv})).
In the notation of (\ref{sbrest}) we have that:
\ba
{}&&( h_{e_J}(1,0) [h_{l_{IJ,\e}}, \tau^j ] )^{A_J}_{\;\;B_J}
\frac{\partial{\hat X}_{j,I}S_{\lambda}(A) }{\partial h_{e_J\;B_J}^{A_J}} = \nonumber\\
&& ( h_{e_J}(1,0) [h_{l_{IJ,\e}}, \tau^j ] )^{B_J}_{\;\;A_J}      (h_{e_I}(1,0)\tau_j)^{B_I}_{\;\;A_I}                                 C_{\lambda}^{A_1..A_N} (\prod_{K\neq I\neq J} h_{e_K}(1,0)^{B_K}{}_{A_K} ) S_{rest}{}_{B_I..B_N}
\ea
If follows from the behavior of holonomies and the interwiner under gauge transformations that under the action of $g$ we have that 
\ba
{}&&( h_{e_J}(1,0) [h_{l_{IJ,\e}}, \tau^j ] )^{A_J}_{\;\;B_J}
\frac{\partial{\hat X}_{j,I}S_{\lambda}(A) }{\partial h_{e_J\;B_J}^{A_J}} \rightarrow \nonumber\\
&&( h_{e_J}(1,0) [h_{l_{IJ,\e}}, g\tau^j g^{-1}] )^{B_J}_{\;\;A_J}      (h_{e_I}(1,0)g\tau_j g^{-1})^{B_I}_{\;\;A_I}                                 
C_{\lambda}^{A_1..A_N} (\prod_{K\neq I\neq J} h_{e_K}(1,0)^{B_K}{}_{A_K} ) S_{rest}{}_{B_I..B_N}\;\;\;\;\;\;\;\;\;\;
\label{3.8}
\ea
Next, note that since the  $j$ index of $\tau_j$ rotates like a  vector index under $SU(2)$ transformations, we have that 
\be
g\tau_j g^{-1} = R_j{}^k \tau_k
\label{taurot}
\ee
where $R_j{}^k$ is an orthogonal matrix so that $R^{jk} R_{jl}=  \delta_l{}^k$. Using the orthogonality property of $R$ together with (\ref{taurot}) in (\ref{3.8}) then implies that the action of the Hamiltonian constraint 
(\ref{3.7}) is gauge invariant.
\\

\noindent (6) Returning to (1) in the light of (2), we note that the precise nature of the loop $l_{IJ,\e}$ depends on how we interpret the action of  $\phi_{I,\e}$ on the vertex structure at $v$. 
In what follows we shall interpret this transformation, as in \cite{p3},
as an `abrupt pulling' of the vertex structure around $v$. 
Hereon till section \ref{sec6} we   restrict attention to  nondegenerate vertices $v$ 
at which no triple of edge tangents are 
linearly dependent. We call such vertices as Grot-Rovelli or `GR' vertices \cite{p3}. For such vertices it is straightforward to implement  $\phi_{I,\e}$ in such a way that no  spurious intersections manifest between
the set of undeformed and deformed edges (for details see \cite{p3}).
This restriction to GR vertices is only for simplicity and we comment on the general case in section \ref{sec6}. 

Let $v$ be a GR vertex (see Figure \ref{undef}).
Since $\phi_{I,\e}$ is the identity outside a coordinate distance of  $O(\e)$  from $v$, the deformation is confined to a small neighbourhood of $v$. From (iii) (see the discussion before (\ref{regqshift})), 
the vertex $v$ is displaced by a coordinate distance $\e$ along the edge $e_I$
to its new position $v_{I,\e}$ at parameter value $t_I=\e_I$. The part of each edge $e_J, J\neq I$ between $v$  (at parameter $t_J=0$) and ${\tilde v}_J$ (at parameter value $t_{\e, I,J}$) is deformed to an edge which connects
$v_{I,\e}$ to ${\tilde v}_J$ so that at the vertex $v_{I,\e}$, the $N-1$ edges $\{e_J, J\neq I\}$ form a `downward pointing cone' with `upward' axis along the $I$th edge, all edges assumed to be outgoing.
Due to this abrupt pulling, a kink is created at each ${\tilde v}_J$. The resulting `conical deformation' is depicted in Figure \ref{condef}.
Since this deformation generates these
kinks it cannot be a {\em diffeomorphism}. This brings us further away from the interpretation of this deformation as a finite transformation generated by the 
smooth vector field ${\hat e}^a_{I,\e}$.  It may be possible to generate such a deformation through an appropriately non-smooth vector field but rather than get into these technical fine points, we feel that at the level of 
heuristics employed in our argumentation, we are justified
in {\em directly} defining the action of $\phi_{I,\e}$ without worrying about whether it can be generated by a vector field.  We note that this sort of an action seems to be exactly the sort of action 
contemplated for the `extended diffeomorphisms' of  \cite{rf}.  Note that from the perspective of curvature approximants this choice of deformation leads to a very natural choice of 
loop $l_{IJ,\e}$. We shall return to this discussion in section \ref{sec6}.
\\

\noindent (7) From (3), it follows that  a second action of the Hamiltonian constraint would not see the new vertices created by the action of the first. As discussed in \cite{jplm} this behavior trivialises the 
constraint algebra. Since we are interested in a non-trivial implementation of the constraint algebra we would like the second action to be non-trivial on deformations created by the first. Such an 
implementation was successfully employed in our demonstration of a non-trivial anomaly free constraint algebra in the toy model context of \cite{p3}. 
In the next section, we improve on the action (\ref{3.7}) so as to generate such  deformations through the addition of higher order terms to  (\ref{3.7}).

\section{\label{sec4}Quantum dynamics through conical deformations}

The reason that the action (\ref{3.6}) results in new vertices which, by virtue of their low valence and planar nature, are expected to be degenerate is that 
the action yields a  {\em sum} over terms each corresponding to the deformation of a {\em single} edge at a time. 
The deformation of the single edge $e_J$ corresponds to its  `abrupt pulling' along the edge $e_I$ so that there are $N-1$ terms in the sum for fixed $I$ in (\ref{3.6}).
In contrast  the desired conical deformation of \cite{p3} involves a deformation of 
all the $N-1$ edges {\em together} along the remaining edge. To achieve this,
the  idea is to convert an expression which looks schematically like 
$\sum_J \delta x_J \frac{\partial F(x_1,..,x_N)}{\partial x_J}$ into  $F( x_1+\delta x_1,.., x_N+ \delta x_N) - F (x_1,..,x_N)$.
In our case $x_J$ corresponds to $h_{e_J}(1,0)$,  $\delta x_J$   to ($h_{l_{IJ,\e}}- {\bf 1}$) and  $F$ to $S_{\lambda}$. Since $F\sim S_{\lambda}$ is a {\em product} (\ref{sbrest}) over the 
$h_{e_J}(1,0)$,  the desired form $F( x_1+\delta x_1,.., x_N+ \delta x_N)$ which accomodates all the deformed holonomies {\em together} is also a product over (deformed) holonomies.  

Hence,  in this section we obtain the desired conical  deformation by converting the {\em sum} over deformations to a difference of {\em products}
over deformations (qualitatively similar to the sum to product modification in \cite{hk}) by the addition of higher order terms to  (\ref{3.6}).
In accordance with the schematics discussed in the previous paragaraph, these terms will be higher order in the (integrated) curvature approximants $h_{l_{IJ,\e}}- {\bf 1}$ 
and hence higher order in the area of the small loops $l_{IJ,\e}$. 
Recall that the group invariance property of the intertwiner (\ref{cgginv}) involves a {\em product} of group elements. This suggests that we attempt 
to  replace the matrix generators of $su(2)$, 
namely the $\tau_j$ matrices, with {\em group} elements and use the group invariance property (\ref{cgginv}) to manipulate   the sum over deformed holonomies in (\ref{3.6}) with the $\tau_j$ so replaced,  into  the desired difference of 
products  which agrees with the sum to leading order. This is the strategy we will follow below. For simplicity we restrict attention to GR vertices i.e. vertices at which no triple of edge tangents is linearly dependent.

Define $g_{j,\delta}$ as the matrix 
\be
g_{j,\delta} = \exp \delta \tau_{j}
\ee
where the spin representation label of $\tau_j$ has been suppressed and will be clear from the context below.
Note that for $\tau_j$ in the $j_J$ representation and $\delta$ small enough (and independent of $\e$), we have that:
\be
\frac{ g^{-1}_{j,\delta} h_{l_{IJ,\e}} g_{j,\delta} - h_{l_{IJ,\e}} }{\delta} =[h_{l_{IJ,\e}} , \tau_j] + O(\delta)
\label{gjdef}
\ee
Using (\ref{gjdef}) we  rewrite (\ref{3.7}) as:
\ba
{\hat H}_{\epsilon}(N) S(A)
&=& \frac{3}{8\pi}N(x(v)) \frac{1}{\e} \sum_{I, J,\;I\neq J } ( h_{e_J}(1,0) [h_{l_{IJ,\e}}, \tau^j ] )^{A_J}_{\;\;B_J}. 
\frac{\partial{\hat X}_{j,I}S_{\lambda}(A) }{\partial h_{e_J\;B_J}^{A_J}} \nonumber\\
&=& 
 \frac{3}{8\pi}N(x(v)) \frac{1}{\e} \sum_{I, J,\;I\neq J } \lim_{\delta\rightarrow 0} \frac{ (h_{e_J}(1,0)g^{-1}_{j,\delta} h_{l_{IJ,\e}} g_{j,\delta} -  h_{e_J}(1,0)h_{l_{IJ,\e}})^{A_J}{}_{B_J}}{\delta}  
\frac{\partial{\hat X}_{j,I}S_{\lambda}(A) }{\partial h_{e_J\;B_J}^{A_J}}\;\;\;\;\;\;\;\;
\label{3.7g} 
\ea
It is useful to define ${\hat H}_{\epsilon, \delta}(N) S(A)$ as
\be
{\hat H}_{\epsilon, \delta}(N) S(A) = 
\frac{3}{8\pi}N(x(v)) \frac{1}{\e \delta} \sum_{I,J,\;I\neq J }   (h_{e_J}(1,0)g^{-1}_{j,\delta} h_{l_{IJ,\e}} g_{j,\delta} -  h_{e_J}(1,0)h_{l_{IJ,\e}}  )^{A_J}_{\;\;B_J}
\frac{\partial{\hat X}_{j,I}S_{\lambda}(A) }{\partial h_{e_J\;B_J}^{A_J}}\;\;\;\;\;\;\;\;
\label{heds}
\ee
Let the area of the small loop $l_{IJ,\e}$ be $\alpha_{I,J,\e}$ and let the largest of these areas for all $I,J$ be $\alpha_{\e}$
We shall proceed  through the addition of terms higher order in $\alpha_{\e}$
\footnote{\label{fnlooparea}While in this work $\alpha_{I,J,\e}, \alpha_{\e} \sim O(\e^2)$,  we prefer to designate the small loop area parameter as  $\alpha_\e$ 
rather than $\e^2$ in anticipation of its role in future work. For a comment in this regard,
please see Remark (4) of section \ref{sec6}.}
to (\ref{heds}) to get our final expression (\ref{4.3})
whose $\delta \rightarrow 0$ limit will be seen to correspond to a modification of (\ref{3.7}) by terms of higher order in $\alpha_{\e}$.

First note that from  (\ref{sbrest}) it follows that:
\be
{\hat X}_{j,I}S_{\lambda} (A) = C_{\lambda}^{A_1..A_N} (\prod_{J\neq I} h_{e_J}(1,0)^{B_J}{}_{A_J} )  ({\hat X}_{j,I}h_{e_I}(1,0))^{B_I}{}_{A_I} S_{rest}{}_{B_I..B_N}(A)
\label{xhatsbrest}
\ee
Next, note that
\be
h_{e_J}(1,0)g^{-1}_{j,\delta} h_{l_{IJ,\e}} g_{j,\delta} -  h_{e_J}(1,0)h_{l_{IJ,\e}} = 
(h_{e_J}(1,0)(g^{-1}_{j,\delta} h_{l_{IJ,\e}} g_{j,\delta} - {\bf 1})) -
 (h_{e_J}(1,0)(h_{l_{IJ,\e}} -{\bf 1}))
\label{hghl-hhl}
 \ee
where each of the two bracketed terms on the right hand side are of the order of the area $\alpha_{\e}$.

It is then straightforward to see that the discussion in the first paragraph of this section together with  the fact that ${\hat X}_{j,I}S_{\lambda}$ is a product over edge holonomies  (\ref{xhatsbrest}), 
and the $O(\alpha_{\e,I,J})$ behavior of the terms on the right hand side of (\ref{hghl-hhl}) implies that:
\ba
{}&&\sum_{J\neq I}h_{e_J}(1,0)(g^{-1}_{j,\delta} h_{l_{IJ,\e}} g_{j,\delta} - {\bf 1}))^{A_J}_{\;\;B_J}\frac{\partial{\hat X}_{j,I}S_{\lambda}(A) }{\partial h_{e_J\;B_J}^{A_J}} 
\nonumber      \\
&=& \left(C_{\lambda}^{A_1..A_N} (\prod_{J\neq I} (h_{e_J}(1,0)g^{-1}_{j,\delta} h_{l_{IJ,\e}} g_{j,\delta})^{B_J}{}_{A_J} )  ({\hat X}_{j,I}h_{e_I}(1,0))^{B_I}{}_{A_I}\; S_{rest}{}_{B_I..B_N}(A)\right)
\nonumber\\
&-& {\hat X}_{j,I}S_{\lambda} (A)  + O(\alpha_{\e}^2),  \label{4.1}
\\
{}&&\sum_{J\neq I }(h_{e_J}(1,0)(h_{l_{IJ,\e}} -{\bf 1}))^{A_J}_{\;\;B_J}\frac{\partial{\hat X}_{j,I}S_{\lambda}(A) }{\partial h_{e_J\;B_J}^{A_J}}
\nonumber \\
&=& \left(C_{\lambda}^{A_1..A_N} (\prod_{J\neq I} (h_{e_J}(1,0) h_{l_{IJ,\e}} )^{B_J}{}_{A_J} )  \; ({\hat X}_{j,I}h_{e_I}(1,0))^{B_I}{}_{A_I}   \; S_{rest}{}_{B_I..B_N}(A)\right)
\nonumber\\
&-& {\hat X}_{j,I}S_{\lambda} (A)  + O(\alpha_{\e}^2) . \label{4.2}
\ea
Using (\ref{4.1}) and (\ref{4.2}) in (\ref{heds}) and dropping the $O(\alpha_\e^2)$ terms, we redefine the action of  ${\hat H}_{\epsilon, \delta}(N)$ to be:
\ba
{}&&{\hat H}_{\epsilon, \delta}(N) S(A) := \frac{3}{8\pi}N(x(v)) \frac{1}{\e \delta} \times \nonumber\\
&&\{\sum_{I=1}^N C_{\lambda}^{A_1..A_N} (\prod_{J\neq I} (h_{e_J}(1,0)g^{-1}_{j,\delta} h_{l_{IJ,\e}} g_{j,\delta})^{B_J}{}_{A_J} )  ({\hat X}_{j,I}h_{e_I}(1,0))^{B_I}{}_{A_I} \; S_{rest}{}_{B_I..B_N}(A)\nonumber \\
&-&\sum_{I=1}^N  C_{\lambda}^{A_1..A_N} (\prod_{J\neq I=1}^N (h_{e_J}(1,0) h_{l_{IJ,\e}} )^{B_J}{}_{A_J} )  \; ({\hat X}_{j,I}h_{e_I}(1,0))^{B_I}{}_{A_I} \; S_{rest}{}_{B_I..B_N}(A)\}
\label{4.3}
\ea
Expanding out the $g^{-1}_{j,\delta}, g_{j,\delta}$ factors in orders of $\delta$, it is straightforward to see that\\
\noindent (a) The term in the second line is the $0$th order in $\delta$  contribution to the term in the first line\\
\noindent (b) The first order in $\delta$ term in the first line can be expanded in powers of the difference between small loop holonomies and the identity i.e. in orders of $\alpha_\e$.
It is straightforward to see that the leading order in $\alpha_\e$ term in the latter expansion 
yields (\ref{3.7}),
and that there are also higher order terms in $\alpha_\e$.
\footnote{{\label{fnorder}} Since (\ref{3.6}) implies that (\ref{3.7}) is of $\frac{O(\alpha_\e)}{\e}$, (\ref{4.3}) is also $\frac{O(\alpha_\e)}{\e}$ to leading order in $\e$ with corrections
of $\frac{O(\alpha_\e^2)}{\e}$.}

It follows from (a), (b) that the $\delta\rightarrow 0$ limit of (\ref{4.3}) yields a higher order in $\alpha_\e$ modification of (\ref{3.7}). As discussed at the beginning of this section, this modification constitutes 
an acceptable definition of the action of ${\hat H}_{\epsilon}(N)$.  Accordingly, 
we define the action of ${\hat H}_{\epsilon}(N)$ on $S(A)$ as:
\be
{\hat H}_{\epsilon}(N) S(A) = \lim_{\delta\rightarrow 0}{\hat H}_{\epsilon, \delta}(N) S(A)
\label{4.4}
\ee
with ${\hat H}_{\epsilon, \delta}(N) S(A)$ defined through (\ref{4.3}). Before taking the $\delta\rightarrow 0$ limit it is useful to
 simplify (\ref{4.3}) by using the gauge invariance properties of the group invariant intertwiner $C_{\lambda}$ (see (\ref{cgginv})). Note that the group invariance property of $C_{\lambda}$ can be rewritten in the form
\be
C_{\lambda}^{A_1..A_N} (\prod_{J\neq I} g^{B_J}{}_{A_J})  = C_{\lambda}^{B_1.. .B_N} g^{-1}{}^{A_I}{}_{B_I}.
\label{cgginv2}
\ee
Using (\ref{deflij}) and setting $g\equiv h_{e_I}(\e_I,0)g_{j,\delta}$ in (\ref{cgginv2}) to simplify the the first part of the second line of (\ref{4.3}), we get: 
\ba
{}&& C_{\lambda}^{A_1..A_N} (\prod_{J\neq I} (h_{e_J}(1,0)g^{-1}_{j,\delta} (h_{l_{IJ,\e}} g_{j,\delta}))^{B_J}{}_{A_J} ) 
\nonumber\\
&=&C_{\lambda}^{D_1..D_N} (\prod_{J\neq I} (h_{e_J}(1,0)g^{-1}_{j,\delta} 
h_{e_J}(t_{\e,J,I}, 0)   )^{-1} h_{\phi_{I,\e}(e_J)}(t_{\e,J,I}, 0)
)^{B_J}{}_{D_J} )  ((h_{e_I}(\e_I,0)g_{j,\delta})^{-1})^{A_I}{}_{D_I})\;\; \;\;\;\;\;\;\;\;
\ea
Using (\ref{deflij}) and setting $g\equiv h_{e_I}(\e_I,0)$ in (\ref{cgginv2}), to simplify the the first part of the third line of (\ref{4.3}), we get: 
\ba
{}&& C_{\lambda}^{A_1..A_N} (\prod_{J\neq I} (h_{e_J}(1,0) h_{l_{IJ,\e}} )^{B_J}{}_{A_J} ) 
 \nonumber\\
&=& C_{\lambda}^{D_1....D_N} (\prod_{J\neq I} (h_{e_J}(1,0)
h_{e_J}(t_{\e,J,I}, 0)   )^{-1} h_{\phi_{I,\e}(e_J)}(t_{\e,J,I}, 0)
)^{B_J}{}_{D_J} ) 
((h_{e_I}(\e_I,0))^{-1})^{A_I}{}_{D_I}\nonumber \\
&=&C_{\lambda}^{D_1...D_N} (\prod_{J\neq I} (h_{e_J}(1,t_{\e,J,I}) h_{\phi_{I,\e}(e_J)}(t_{\e,J,I}, 0)
)^{B_J}{}_{D_J} )   ((h_{e_I}(\e_I,0))^{-1})^{A_I}{}_{D_I}
\ea
Using these in (\ref{4.3}), we get
\ba
&&{\hat H}_{\epsilon, \delta}(N) S(A) := \frac{3}{8\pi}N(x(v)) \frac{1}{\e \delta}  S_{rest}{}_{B_I..B_N}(A)) C_{\lambda}^{D_1..D_N}       \nonumber\\
&\big(&\sum_{I=1}^N (\prod_{J\neq I} (h_{e_J}(1,0)g^{-1}_{j,\delta} 
(h_{e_J}(t_{\e,J,I}, 0)   )^{-1} h_{\phi_{I,\e}(e_J)}(t_{\e,J,I}, 0)
)^{B_J}{}_{D_J} ) 
((h_{e_I}(\e_I,0)g_{j,\delta})^{-1})^{A_I}{}_{D_I}
({\hat X}_{j,I}h_{e_I}(1,0))^{B_I}{}_{A_I}\nonumber \\
&-&\sum_{I=1}^N (\prod_{J\neq I} (h_{e_J}(1,t_{\e,J,I}) h_{\phi_{I,\e}(e_J)}(t_{\e,J,I}, 0)
)^{B_J}{}_{D_J} )  
((h_{e_I}(\e_I,0))^{-1})^{A_I}{}_{D_I}
({\hat X}_{j,I}h_{e_I}(1,0))^{B_I}{}_{A_I}  \;\;   \big) 
\label{4.5}
\ea
Taking the $\delta \rightarrow 0$ limit of (\ref{4.5}), we obtain:
\ba
&&{\hat H}_{\epsilon}(N) S(A) := -\frac{3}{8\pi}N(x(v)) \frac{1}{\e}  S_{rest}{}_{B_I..B_N}(A)) C_{\lambda}^{D_1..D_N}  \nonumber\\
&\big(& \sum_{I=1}^N  \{ \sum_{J\neq I} (\prod_{K\neq I,J} h_{e_K}(1,t_{\e,K,I}) h_{\phi_{I,\e}(e_K)}(t_{\e,K,I}, 0)
)^{B_K}{}_{D_K} )      \nonumber\\
&&(h_{e_J}(1,0)\tau_j 
(h_{e_J}(t_{\e,J,I}, 0)   )^{-1} h_{\phi_{I,\e}(e_J)}(t_{\e,J,I}, 0)
)^{B_J}{}_{D_J} 
((h_{e_I}(\e_I,0))^{-1})^{A_I}{}_{D_I} 
({\hat X}_{j,I}h_{e_I}(1,0))^{B_I}{}_{A_I} \}   \nonumber \\
&+& \sum_{I=1}^N\{ (\prod_{J\neq I} (h_{e_J}(1,t_{\e,J,I}) h_{\phi_{I,\e}(e_J)}(t_{\e,J,I}, 0)
)^{B_J}{}_{D_J} )
(\tau_j(h_{e_I}(\e_I,0))^{-1})^{A_I}{}_{D_I}
({\hat X}_{j,I}h_{e_I}(1,0))^{B_I}{}_{A_I}\}\;\big)
\;\;\;\;\;\;\;\;\;\;\; 
\label{4.6}
\ea
Note that in the last line of (\ref{4.6}):
\ba
(\tau_j(h_{e_I}(\e_I,0))^{-1})^{A_I}{}_{D_I}
({\hat X}_{j,I}h_{e_I}(1,0))^{B_I}{}_{A_I}
&=&(\tau_j(h_{e_I}(\e_I,0))^{-1})^{A_I}{}_{D_I}
(h_{e_I}(1,0)\tau_j)^{B_I}{}_{A_I}\nonumber\\
= 
(h_{e_I}(1,0)\tau_j\tau_j(h_{e_I}(\e_I,0))^{-1})^{B_I}{}_{D_I}
&=&  -j_I(j_I+1)  h_{e_I}(1, \e_I)^{B_I}{}_{D_I}
\label{4.7}
\ea
where in the last line we have used the fact that $\tau_j$ is the  spin $j_I$ representative of the $j$th generator of $SU(2)$. 

Using (\ref{4.7}) we have the result:
\ba
&&{\hat H}_{\epsilon}(N) S(A) := \frac{3}{8\pi}N(x(v)) \frac{1}{\e}  S_{rest}{}_{B_I..B_N}(A) C_{\lambda}^{D_1..  .D_N}\nonumber\\
&\big(&\sum_{I=1}^N   \{ (j_I)(j_I+1)  (\prod_{J\neq I} (h_{e_J}(1,t_{\e,J,I}) h_{\phi_{I,\e}(e_J)}(t_{\e,J,I}, 0)
)^{B_J}{}_{D_J} )  h_{e_I}(1,\e_I)^{B_I}{}_{D_I} \}
\;\;\;\;\;\;\;\;\;\;\; \nonumber \\
&-&\sum_{I=1}^N  \{ \sum_{J\neq I} (\prod_{K\neq I,J } h_{e_K}(1,t_{\e,K,I}) h_{\phi_{I,\e}(e_K)}(t_{\e,K,I}, 0)
)^{B_K}{}_{D_K} )      \nonumber\\
&&(h_{e_J}(1,0)\tau_j 
(h_{e_J}(t_{\e,J,I}, 0)   )^{-1} h_{\phi_{I,\e}(e_J)}(t_{\e,J,I}, 0)
)^{B_J}{}_{D_J} 
(h_{e_I}(1,0)\tau_j)^{B_I}{}_{A_I}
((h_{e_I}(\e_I,0))^{-1})^{A_I}{}_{D_I} \}
\; \big)\;\;\;\;\;\;\;\;\;\;\;
\label{4.8}
\ea
We write the action (\ref{4.8}) in the concise form:
\be 
{\hat H}_{\epsilon}(N) S(A) := \frac{3}{8\pi}N(x(v)) \frac{\sum_{I=1}^N j_I(j_I+1)S_{\lambda,I, \e} - \sum_{I=1}^N\sum_{J\neq I} S_{\lambda,I,J,\e}}{\e}
\label{4.8a}
\ee
where we have defined 
the deformed states $S_{\lambda,I, \e}, S_{\lambda,I,J,\e}$ to be:
\be
S_{\lambda,I, \e} := C_{\lambda}^{D_1..  .D_N} (\prod_{J\neq I} (h_{e_J}(1,t_{\e,J,I}) h_{\phi_{I,\e}(e_J)}(t_{\e,J,I}, 0)
)^{B_J}{}_{D_J} )  h_{e_I}(1,\e_I)^{B_I}{}_{D_I} S_{rest\;B_1..B_N},
\label{sie}
\ee
\ba
&&S_{\lambda,I,J,\e} := 
C_{\lambda}^{D_1..D_N} (\prod_{K\neq I,J } h_{e_K}(1,t_{\e,K,I}) h_{\phi_{I,\e}(e_K)}(t_{\e,K,I}, 0)
)^{B_K}{}_{D_K} )      \nonumber\\
&&(h_{e_J}(1,0)\tau_j 
(h_{e_J}(t_{\e,J,I}, 0)   )^{-1} h_{\phi_{I,\e}(e_J)}(t_{\e,J,I}, 0)
)^{B_J}{}_{D_J} 
(h_{e_I}(1,0)\tau_j)^{B_I}{}_{A_I}
((h_{e_I}(\e_I,0))^{-1})^{A_I}{}_{D_I} S_{rest\;B_1..B_N}. \;\;\;\;\;\;\;\;\;\;\;\;\;\;\;\;
\label{sije}
\ea
Equation (\ref{4.8a}) is our final result. We make the following remarks.\\

\begin{figure}[H]
\centering
  \begin{subfigure}[H]{0.3\textwidth}
    \centering
    \includegraphics[width=\textwidth]{Figura2.pdf}
    \caption{}
 \label{figconicala}
  \end{subfigure} \quad
  \begin{subfigure}[H]{0.3\textwidth}
  \centering 
    \includegraphics[width=\textwidth]{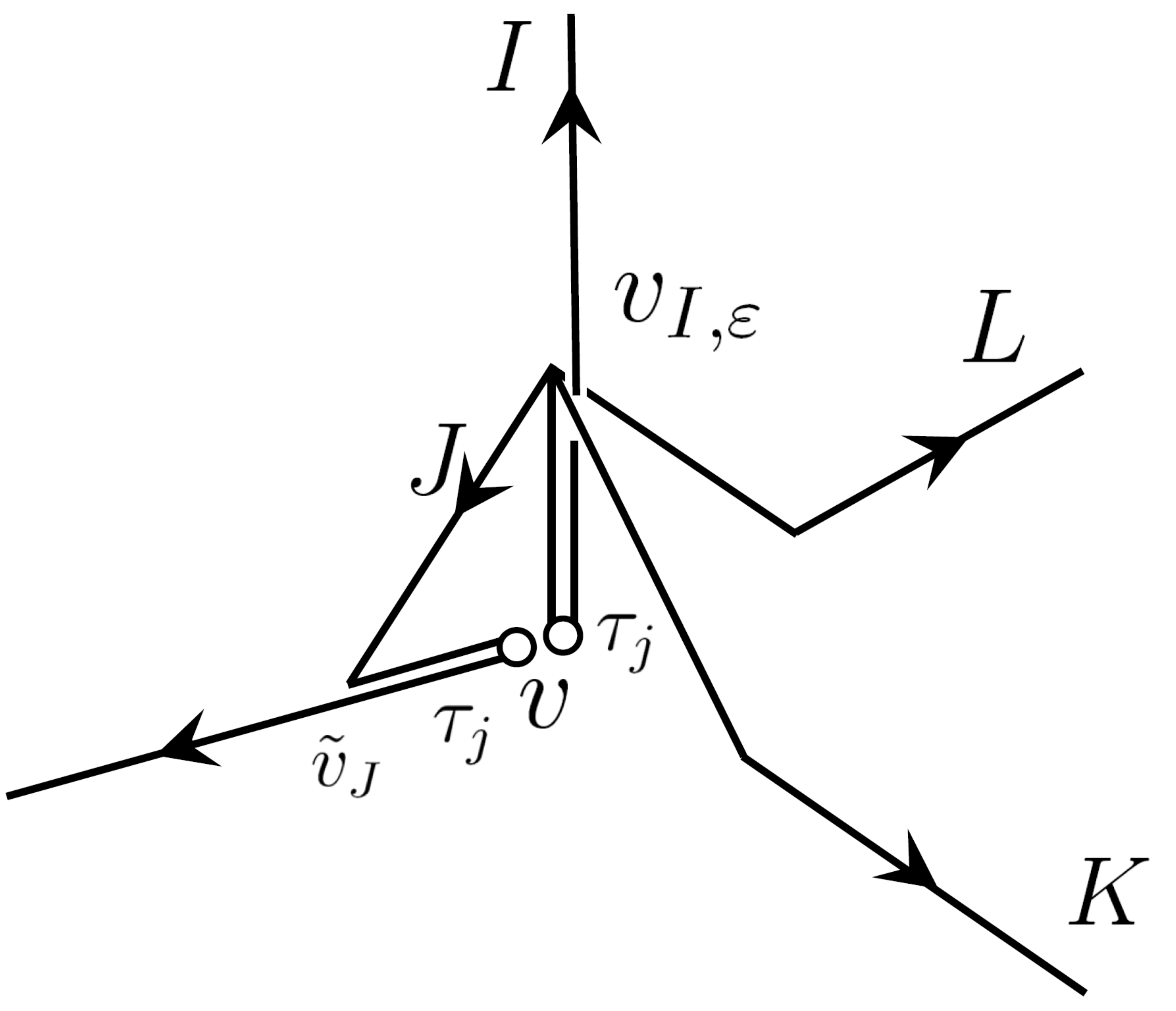}
    \caption{}
   \label{figconicalb}
  \end{subfigure} 
  \caption{ The structures of the conical deformed states $S_{\lambda, I, \e}, S_{\lambda, I, J, \e}$ near the vertex $v$ of $S$ are depicted above. Figure \ref{figconicala} is identical to Figure \ref{condef} and
  depicts the vertex structure of  $S_{\lambda, I, \e}$. Figure \ref{figconicalb} depicts  the vertex structure of $S_{\lambda, I, J, \e}$.  The intertwiner $C_{\lambda}$ sits at the displaced vertex $v_{I,\e}$. 
The circles in Figure \ref{figconicalb} depict insertions of $\tau_j$
 at the  original vertex $v$. The $I$th edge at $v_{I,\e}$ starts out oriented opposite to its counterpart in the undeformed state $S$ (see Figure \ref{undef}) till $v$ where there is an insertion of $\tau_j$. It then 
traces out the same trajectory as its counterpart in $S$.  The insertion structure on the undeformed part of  $J$th edge  between ${\tilde v}_J$ and $v$  is similarly depicted: the segment starts from ${\tilde v}_J$ moves
opposite to its counterpart in $S$ to $v$ where there is an insertion and then runs back.
  }
\label{figconical}%
\end{figure}

\noindent (1) {\em Conically deformed states:} 
 $S_{\lambda,I, \e},\; S_{\lambda,I,J,\e}$ correspond to states in which  the vertex deformations in the vicinity of the vertex $v$ which are defined through (\ref{sie}), (\ref{sije}).
As shown in Figure \ref{figconical}, both (\ref{sie}) and  (\ref{sije}) describe deformations which create a new 
conically deformed vertex of the type encountered in the $U(1)^3$ toy model in  \cite{p3}. Figure \ref{figconicala} depicts the `electric diffeomorphism' type deformation \cite{p3} described by (\ref{sie}) wherein the 
original vertex at $v$  is displaced to its new position along the $I$th edge and the edge tangents at $v$ are conically deformed to yield the vertex structure at the displaced vertex, resulting in the state $S_{\lambda,I, \e}$
In Figure \ref{figconicalb} the state $S_{\lambda,I,J,\e}$ described by (\ref{sije}) 
has both the new conically deformed vertex
as well as the original one except that the valence of the original vertex has decreased to 2 and the original vertex is therefore degenerate. The second vertex is actually $N+1$ valent but using the terminology of \cite{p3} 
we shall refer to it as an $N$ valent singly CGR vertex. A CGR vertex is one in which a pair of edge tangents are collinear and which is GR with respect to the set of edge tangents obtained by dropping one
of this collinear pair. The  state $S_{\lambda,I,J,\e}$ is not itself a spin network state because the initial parts of the $I,J$ edges are retraced after insertions of $\tau_i$.  However it can of course be decomposed into spin network states.
The Lemma in Appendix \ref{seca2.1} shows that with an appropriate choice of intertwiners,  the result  can be written as {\em single} spin network with each of these parts labelled by spin 1.
\\

\noindent (2) {\em Gauge Invariance}:
It is straightforward to see, using an analysis similar to that of Remark (5) at the end of  section \ref{sec3}, that (\ref{sie}), (\ref{sije}) are   gauge invariant expressions.
The gauge invariance of $S_{\lambda,I, \e}$ may also be inferred directly from Figure \ref{figconicala}:  the figure shows that 
  $S_{\lambda,I, \e}$ is obtained from  $S_{\lambda}$ by a displacement of the vertex structure near $v$ in $S_{\lambda}$ by $\phi_{I,\e}$  thus implying that the   gauge invariance of $S_{\lambda,I, \e}$
  follows immediately from that of $S_{\lambda}$. For $S_{\lambda, I, J, \e}$ gauge invariance under gauge transformations supported away from $v_{I,\e}, v$ is immediate. For gauge transformations at $v_{I,\e}$ gauge invariance
  follows from group invariance of $C_{\lambda}$. For gauge transformations at $v$ gauge invariance follows from an argument identical to that in Remark (5) of section 3 based on (\ref{taurot}).
\\

\noindent (3) {\em Constraint action and deformed states in orders of $\alpha_e$}:
From Footnote \ref{fnorder}, the combination of the two terms in the curly brackets of (\ref{4.8}) is of order $\alpha_{\e}$. Consequently the right hand side of (\ref{4.8a}) is also of
order $\alpha_{\e}$. It is useful for our considerations in section \ref{sec5} to display the constraint action in a form which explicitly segregates the order $\alpha_{\e}$ contributions to $S_{\lambda,I, \e}, S_{\lambda,I,J,\e}$.
The relevant computations are straightforward and we relegate them to Appendix \ref{seca1}. The desired form of (\ref{4.8a}) turns out to be:
\be 
{\hat H}_{\epsilon}(N) S(A) := \frac{3}{8\pi}N(x(v)) \big(\sum_{I=1}^N\frac{\{ j_I(j_I+1)(S_{\lambda,I, \e} - S_{\lambda})\}}{\e} - \sum_{I=1}^N\frac{\{(\sum_{J\neq I} S_{\lambda,I,J,\e})  - j_I(j_I+1)S_{\lambda}\}}{\e}\big).
\label{4.14}
\ee
As shown in the Appendix, each of the curly brackets in (\ref{4.14})   is  of $O(\alpha_\e)$.  Note that 
that equation (\ref{4.14}) is exactly the same as (\ref{4.8}) because we have only added and subtracted the $S_{\lambda}(A)$  term.
The form of the constraint action (\ref{4.14}) will be of use in section \ref{sec5} wherein  we further modify the constraint action by terms which are higher order in $\alpha_\e$.
\\


\noindent (4) {\em Comments on  anomaly free-ness and propagation}:
The form of the constraint action (\ref{4.14}) is 
reminiscent of the ``${\hat U}-{\bf 1}$'' form  in the $U(1)^3$ toy model \cite{p3}. This ``${\hat U}-{\bf 1}$'' form played a crucial role 
in the demonstration of the anomaly free property of the constraint action in \cite{p3}. 
Therefore one may try to repeat the analysis of \cite{p3} and attempt to show that the constraint action (\ref{4.14}) is consistent with an anomaly free constraint algebra.
Before doing so, it is of interest to  modify (\ref{4.14})  through the addition of higher order terms so as to 
bring the constraint action into an even  more optimal form to which the general methodology  of \cite{p3} could be applied to investigate the issue of anomaly free-ness.
We do this in the Appendix \ref{seca2}. Since the modified  deformations have no significant qualitative difference with those generated by the action (\ref{4.14}),
we do not display the modified constraint action here; it is of potential interest only for attempts at tackling the anomaly free issue along the lines of \cite{p3}.

As mentioned in section \ref{sec1}, ongoing preliminary efforts to apply the considerations of \cite{p3}, developed in  the $U(1)^3$ setting, to the $SU(2)$ case of interest here meet with 
significant technical complications. A brief description of these complications follows and may be skipped on a first reading. As mentioned in (1) above, the action of the constraint
on the GR vertex $v$ creates the states $S_{\lambda, I,J,\e}$  with a CGR vertex. The action of the constraint can be generalised in a straightforward manner so as to act on this CGR vertex,
It then turns out that (a) among the set of deformed states generated by  repeated actions of the constraint, there is a subset of states with vertices of increasing valence and (b) these states
complicate attempts to demonstrate the existence of an anomaly free constraint algebra.
In \cite{p3}, this phenomenon of increasing valence was negated by a more involved choice of constraint action on CGR vertices through the use of technical tools called {\em interventions}.
A generalisation of the application of interventions to the $SU(2)$ case seems to lead  to a very complicated and baroque action for the constraint. 
We also note that the action discussed in \cite{p3}, while anomaly free in the specific sense defined in that work,  is not consistent with 3d propagation \cite{u13prop}. It is then likely that any  
intervention based generalisation of the considerations of 
\cite{p3} to the $SU(2)$ case would also suffer from this lack of adequate propagation. 
Hence in section \ref{sec5}  we further modify  the action (\ref{4.14})  through higher order terms so as to obtain a constraint action which, we believe, 
stands a better chance of displaying consistency with the properties of an anomaly free constraint algebra and  propagation {\em together}.
\\

\section{\label{sec5}Mixed action quantum dynamics}

In section \ref{sec5.1} we define a new constraint action by modifying  the second set of contributions  in (\ref{4.14})  through the addition of higher order terms in $\alpha_\e$ to $S_{\lambda, I,J,\e}$.
The resulting set of deformed spin nets are expected to facilitate propagation for reasons which we discuss in section \ref{sec6}. 
In section \ref{sec5.2} we compute the commutator of two such Hamiltonian constraint  actions. We also construct a particular operator correspondent of the Poisson bracket between the corresponding pair of 
classical Hamiltonian constraints. We find that the deformed spin nets generated by this operator correspondent of the constraint Poisson bracket are closely related  to those
generated by the constraint operator  commutator. We  discuss  this property in relation to that of an uncoventional view of the anomaly free property in section \ref{sec5.3}. As in section \ref{sec4} we restrict attention to
the constraint action on GR vertices i.e. vertices at which no triplet of edge tangents is linearly dependent. Additionally, in sections \ref{sec5.2} and \ref{sec5.3} we shall use density weight one 
constraints.
\subsection{\label{sec5.1}The Hamiltonian constraint action}
Starting from   (\ref{sije}), we have that:
\ba
&&S_{\lambda,I,J,\e} := 
C_{\lambda}^{D_1..D_N} (\prod_{K\neq I,J } h_{e_K}(1,0) h_{l_{IK,\e}}(h_{e_I}(\e,0))^{-1}
)^{B_K}{}_{D_K} )      \nonumber\\
&&(h_{e_J}(1,0)\tau_j 
h_{l_{IJ,\e}}(h_{e_I}(\e,0))^{-1}
)^{B_J}{}_{D_J} 
(h_{e_I}(1,0)\tau_j)^{B_I}{}_{A_I}
((h_{e_I}(\e_I,0))^{-1})^{A_I}{}_{D_I} S_{rest\;B_1..B_N} \;\;\;\;\;\;\;\;
\nonumber\\
&&:= 
C_{\lambda}^{D_1..D_N} (\prod_{K\neq I,J } (h_{e_K}(1,0) h_{l_{IK,\e}}
)^{B_K}{}_{D_K} )
(h_{e_J}(1,0)\tau_j 
h_{l_{IJ,\e}}
)^{B_J}{}_{D_J} 
(h_{e_I}(1,0)\tau_j)^{B_I}{}_{D_I}
 S_{rest\;B_1..B_N}, \;\;\;\;\;\;\;\;
\label{5.1}
 \ea
where we have used (\ref{deflij}) in the first equation   and (\ref{cgginv}) in the second. Next we set \\$h_{l_{IM,\e}}= (h_{l_{IM,\e}} -{\bf 1}) +{\bf 1}$ and use the fact that 
$h_{l_{IM,\e}} -{\bf 1} = O(\alpha_\e)$ to expand (\ref{5.1}) in   $\alpha_\e$. We get:
\ba
&&S_{\lambda,I,J,\e} := \{S_{rest\;B_1..B_N} C_{\lambda}^{D_1..D_N} \nonumber\\
&& (\prod_{K\neq I,J } h_{e_K}(1,0) 
^{B_K}{}_{D_K} )      
(h_{e_J}(1,0)\tau_j 
)^{B_J}{}_{D_J} 
(h_{e_I}(1,0)\tau_j)^{B_I}{}_{D_I}
\nonumber\\
& +&\sum_{K\neq I,J }(\prod_{M\neq K\neq J\neq I} h_{e_M}(1,0)^{B_M}{}_{D_M})  (h_{e_K}(1,0)(h_{l_{IK,\e}} -{\bf 1})
)^{B_K}{}_{D_K} )     
(h_{e_J}(1,0)\tau_j 
)^{B_J}{}_{D_J} 
(h_{e_I}(1,0)\tau_j)^{B_I}{}_{D_I} \nonumber\\
&+& (\prod_{K\neq I,J } h_{e_K}(1,0) 
^{B_K}{}_{D_K} )      
(h_{e_J}(1,0)\tau_j 
(h_{l_{IJ,\e}} -{\bf 1})
)^{B_J}{}_{D_J} 
(h_{e_I}(1,0)\tau_j)^{B_I}{}_{D_I}\} + O(\alpha_\e^2)
\label{5.2}
\ea
We shall define a new Hamiltonian constraint action from (\ref{4.14})  by neglecting the  $O(\alpha_\e^2)$ contribution to $S_{\lambda,I,J,\e}$ in (\ref{5.2}). It is convenient to abuse notation slightly
and continue to denote the right hand side of (\ref{5.2}) without the $O(\alpha_\e^2)$ term by $S_{\lambda,I,J,\e}$.
Using this notation, it is straightforward to see that (\ref{5.2}) (without the $O(\alpha_\e^2)$ term) can be expanded to give:
\ba
S_{\lambda,I,J,\e} &:=& S_{rest\;B_1..B_N} C_{\lambda}^{D_1..D_N} \nonumber\\
&\big(& -(N-2) (\prod_{K\neq I,J } h_{e_K}(1,0) 
^{B_K}{}_{D_K} )      
(h_{e_J}(1,0)\tau_j 
)^{B_J}{}_{D_J} 
(h_{e_I}(1,0)\tau_j)^{B_I}{}_{D_I}
\nonumber\\
& +&\sum_{K\neq I,J }(\prod_{M\neq K\neq J\neq I} h_{e_M}(1,0)^{B_M}{}_{D_M})  (h_{e_K}(1,0)h_{l_{IK,\e}}
)^{B_K}{}_{D_K} )     
(h_{e_J}(1,0)\tau_j 
)^{B_J}{}_{D_J} 
(h_{e_I}(1,0)\tau_j)^{B_I}{}_{D_I}\nonumber\\
&+& (\prod_{K\neq I,J } h_{e_K}(1,0) 
^{B_K}{}_{D_K} )      
(h_{e_J}(1,0)\tau_j 
h_{l_{IJ,\e}} 
)^{B_J}{}_{D_J} 
(h_{e_I}(1,0)\tau_j)^{B_I}{}_{D_I}  \;\;\big)
\label{5.3}
\ea
The new constraint action is then defined to be:
\be 
{\hat H}_{\epsilon}(N) S(A) := \frac{3}{8\pi}N(x(v)) \big(\sum_{I=1}^N\frac{\{ j_I(j_I+1)(S_{\lambda,I, \e} - S_{\lambda})\}}{\e} - \sum_{I=1}^N\frac{\{\sum_{J\neq I} S_{\lambda,I,J,\e}  - j_I(j_I+1)S_{\lambda}\}}{\e}\big).
\label{5.4}
\ee
with 
$S_{\lambda,I,J,\e}$ given by (\ref{5.3}) rather than (\ref{sije}).

Next, note that the infinitesmal form of the group invariance property (\ref{cgginv}) implies that:
\be
\sum_{I=1}^N C_{\lambda}^{B_1..B_{I-1}A_IB_{I+1}...B_N} \tau_j^{B_I}{}_{A_I} =0
\label{infgginv}
\ee
Using (\ref{infgginv}) to simplify the sum over $I\neq J$ of the first line of (\ref{5.3}), we obtain:
\ba
&&-(N-2)\sum_{J\neq I}S_{rest\;B_1..B_N} C_{\lambda}^{D_1..D_N}
  (\prod_{K\neq I,J } h_{e_K}(1,0) 
^{B_K}{}_{D_K} )      
(h_{e_J}(1,0)\tau_j 
)^{B_J}{}_{D_J} 
(h_{e_I}(1,0)\tau_j)^{B_I}{}_{D_I} \nonumber\\
&=&
(N-2) S_{rest\;B_1..B_N} C_{\lambda}^{D_1..D_N} 
 (\prod_{K\neq I} h_{e_K}(1,0) 
^{B_K}{}_{D_K} )      
(h_{e_I}(1,0)\tau_j\tau_j)^{B_I}{}_{D_I}) \nonumber \\
&=& -(N-2) (j_I(j_I+1)) S_{\lambda}
\label{5.5}
\ea
Defining:
\ba
&&S_{\lambda,(1)I,J,K,\e} =S_{rest\;B_1..B_N} C_{\lambda}^{D_1..D_N} \nonumber\\
&&(\prod_{M\neq K\neq J\neq I} h_{e_M}(1,0)^{B_M}{}_{D_M})  (h_{e_K}(1,0)h_{l_{IK,\e}}
)^{B_K}{}_{D_K}      
(h_{e_J}(1,0)\tau_j 
)^{B_J}{}_{D_J} 
(h_{e_I}(1,0)\tau_j)^{B_I}{}_{D_I} ,\;\;\;\;\;\;
\label{s1}
\ea
\be
S_{\lambda,(2)I,J,\e} =S_{rest\;B_1..B_N} C_{\lambda}^{D_1..D_N} 
(\prod_{K\neq I,J } 
h_{e_K}(1,0) 
^{B_K}{}_{D_K} )      
(h_{e_J}(1,0)\tau_j 
h_{l_{IJ,\e}} 
)^{B_J}{}_{D_J} 
(h_{e_I}(1,0)\tau_j)^{B_I}{}_{D_I}
\label{s2}
\ee
we have, finally:
\ba
{\hat H}_{\epsilon}(N) S(A) &:= &\frac{3}{8\pi}N(x(v)) \big(\;\sum_{I=1}^N\frac{ j_I(j_I+1)(S_{\lambda,I, \e} - S_{\lambda})}{\e}
\nonumber\\ 
&-& \sum_{I=1}^N\frac{  (\sum_{J\neq I}\sum_{K\neq I,J}S_{\lambda,(1)I,J,K,\e}) +(\sum_{J\neq I} S_{\lambda,(2)I,J,\e}) - j_I(j_I+1)(N-1)S_{\lambda}}{\e}\;\big).\;\;\;\;\;\;\;\;\;\;
\label{5.6}
\ea

We make the following remarks:
\\

\begin{figure}[H]
\centering
  \begin{subfigure}[H]{0.3\textwidth}
    \centering
    \includegraphics[width=\textwidth]{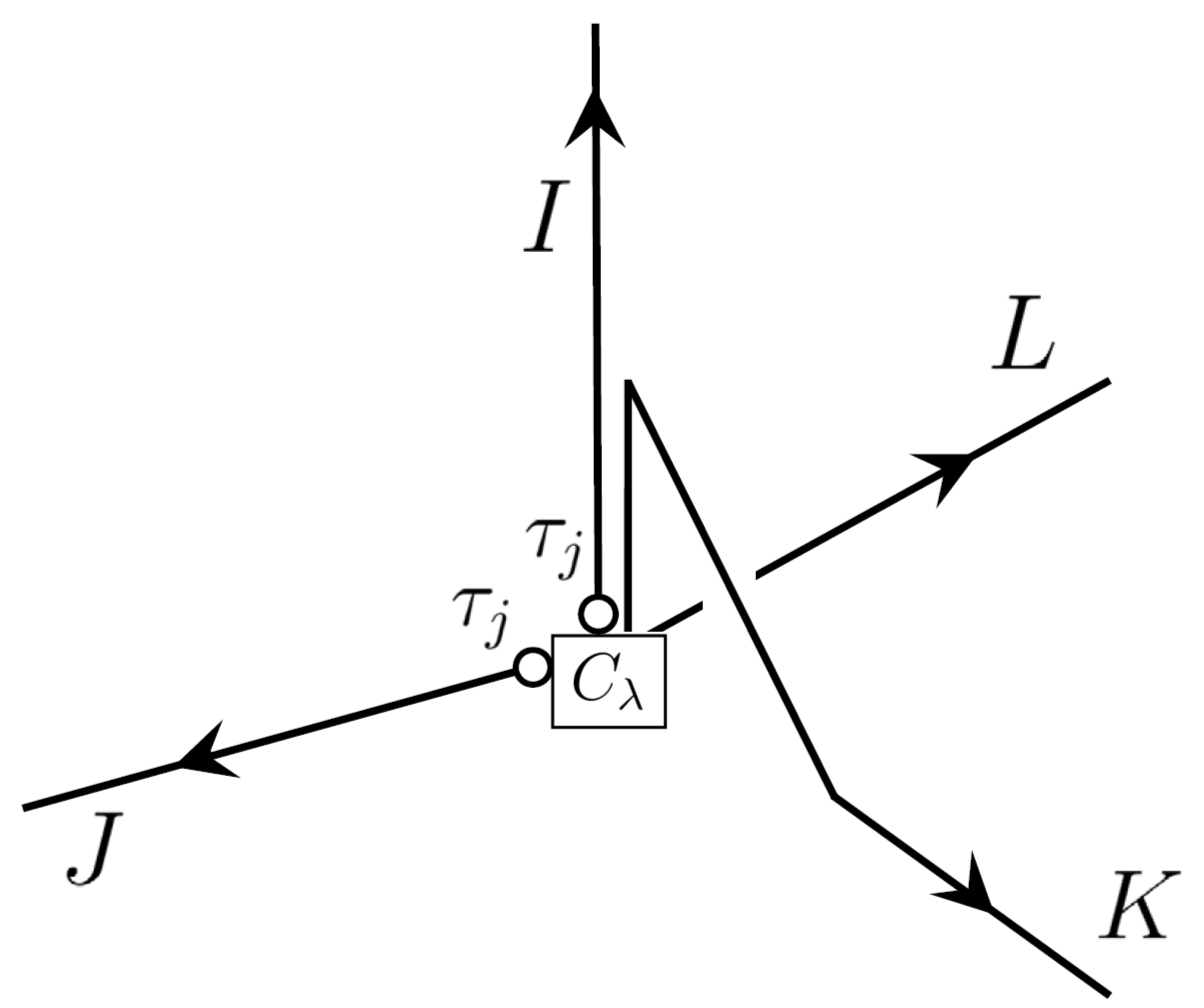}
    \caption{}
 \label{figmixed1}
  \end{subfigure} \quad
  \begin{subfigure}[H]{0.3\textwidth}
  \centering 
    \includegraphics[width=\textwidth]{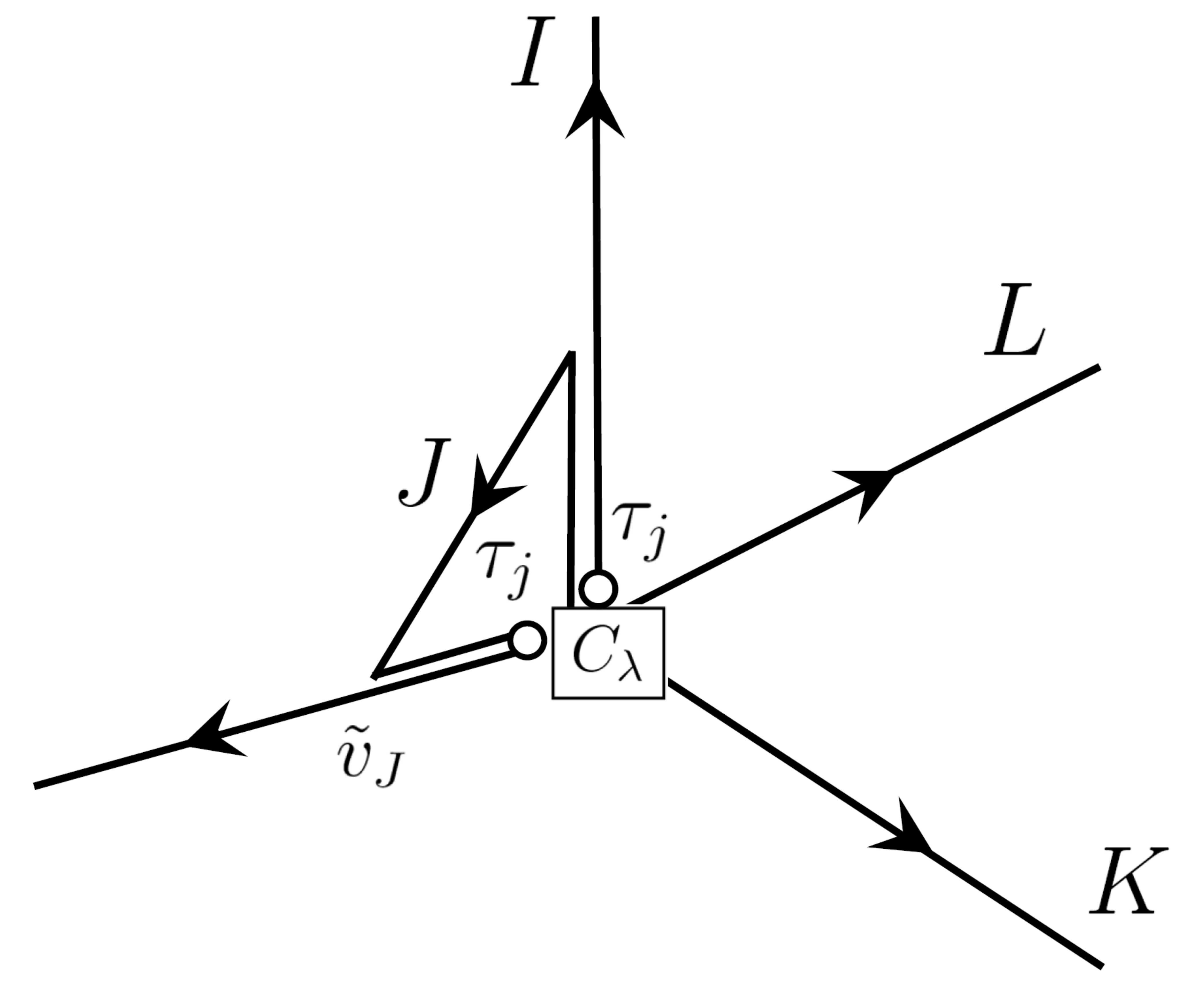}
    \caption{}
   \label{figmixed2}
  \end{subfigure} 
  \caption{ 
Figures \ref{figmixed1}, \ref{figmixed2}  depict the deformed vertex structure of $S_{\lambda,(1)I,J,K,\e},S_{\lambda,(2)I,J,\e} $ respectively. The box represents the intertwiner $C_{\lambda}$ located at the 
position of the undisplaced vertex $v$ of $S$ (see Figure \ref{undef}). Similar to Figure \ref{figconical} insertions of $\tau_j$ are represented by circles. The touching of an object with the interwtiner box indicates
an index contraction between an index of the object and that of the intertwiner so that whereas both the $\tau_j$'s have index contractions with $C_{\lambda}$ in Figure \ref{figmixed1}, in Figure \ref{figmixed2} only the 
one on the $I$th edge has such an index contraction.
}
 \label{figmixed}%
\end{figure}

\noindent (1) 
We depict the vertex deformations in $S_{\lambda,(1)I,J,K,\e},S_{\lambda,(2)I,J,\e}$  in Fig \ref{figmixed}.
The displaced vertex in each of the states $S_{\lambda,(1)I,J,K,\e},S_{\lambda,(2)I,J,\e}$ is planar with valence at most 3.
The kinks which are created by the deformation are also planar and at most trivalent. The states $S_{\lambda,(1)I,J,K,\e},S_{\lambda,(2)I,J,\e}$  can be expanded in terms of spin networks.
It follows that the displaced vertex and the kinks created by the deformation  in each of these spin networks  are also planar, at most trivalent and, from (2) below,  gauge invariant.
Hence these  vertices have vanishing volume.
The inverse metric determinant whether defined through a Thiemann like
trick \cite{qsd} with the Ashtekar-Lewandowski volume operator \cite{alvolume} or a Tikhonov regularization (See Footnote \ref{fntycho}), vanishes and these kinks and the displaced vertex are {\em degenerate} so that a second constraint action does not see them. 

\noindent (2) Gauge invariance of the states $S_{\lambda,(1)I,J,K,\e},S_{\lambda,(2)I,J,\e}$  follows  immediately from an  argumentation along the lines of that employed in  Remark (5), section \ref{sec3.3}.
\\

\noindent (3) 
The Lemma in the Appendix \ref{seca2.1} may be applied to the retraced part of the $J$th edge in $S_{\lambda,(2)I,J,\e}$
with the implication that every spin network in the spin network decomposition of $S_{\lambda,(2)I,J,\e}$ has this part of the $J$th edge labelled by spin 1.
The initial part of the $I$th edge of a spin network in the spin network decomposition of  $S_{\lambda,(2)I,J,\e}$ can carry a  spin label $j$  in the range  $|j_J-j_I|$ to $j_J+j_I$ in accordance with the 
Clebsch- Gordan decomposition. 
If $j=0$ is permissible, the corresponding spin network has an $N-1$ valence vertex at $v$. 
Spin nets with $j\neq 0$ have an $N$ valent vertex at $v$. For $S_{\lambda,(1)I,J,K,\e}$ similar conclusions hold except that $J$ is replaced by $K$, $N$ by $N-1$, and $N-1$ by $N-2$.
If the vertex $v$ is non-degenerate it can support a second action of the Hamiltonian constraint but this action does not contribute to the commutator as it acts at the same vertex as the first constraint.
\\


\subsection{\label{sec5.2} A curious property of the commutator}
In this section and the next, we shall consider unit density weight constraints. As indicated in Remark (3) of section \ref{sec3.3}, the considerations of that section go through for any density weight of the Hamiltonian constraint.
It is straightforward to check that this holds for our argumentation in sections \ref{sec4} and \ref{sec5.1} as well, the only change being in the overall power of $\e$ in the expression for the constraint action and
a possible change in the intertwiner $C_{\lambda}$ which is obtained  from the intertwiner $C$ of the state $S$   through the action of the  power of the inverse metric determinant operator which governs the density weight of the 
constraint.  In particular, for the case of density weight 1, all explicit factors of $\e$ cancel so that there is not such factor at all, consistent with the observations of \cite{qsd}.
Moreover, the lapses are scalars and no coordinate patch is required for their evaluation.

We first compute the commutator between two constraint actions, each given by (\ref{5.6}) i.e. we compute, with $\e_1<<\e$, the object:
\be
[{\hat H}(M),{\hat H}(N)]_{\e_1, \e} S(A) := 
({\hat H}_{\epsilon_1}(M){\hat H}_{\epsilon}(N)-{\hat H}_{\e_1}(N){\hat H}_{\epsilon}(M)) S(A)
\ee
For simplicity we assume that $N,M$ are such that the only nondegenerate vertex in their support is $v$. Terms which involve the evaluation of both lapses at $v$ vanish by antisymmetry. It follows that the contributions from the 
commutator come from the action of a second constraint on deformed states of the type  $S_{\lambda,I, \e}$ created by the first:
\ba 
&&[{\hat H}(M),{\hat H}(N)]_{\e_1, \e} S =(\frac{3}{8\pi})^2 \sum_I j_I(j_I+1) (N(v) M(v_{I,\e})- M(v)N(v_{I,\e}))\times \;\;\;\;\;\;\;\;\;\;\;\;\;\;\;\;\;
\;\;\;\;\;\;\;\;\;\;\;\;\;\;\;\;\;\;\;\;\;\;\;\;\;\;\;\;\;\;\;\;\;
\nonumber \\
&\big( &\sum_{I_1} j_{I_1}(j_{I_1} +1) (S_{[(\lambda^I, I_1, \e_1),(\lambda,I, \e)]} - S_{\lambda, I,\e})-  \sum_{I_1}\sum_{J_1\neq I_1}\sum_{K_1\neq I_1,J_1}S_{ [(\lambda^I,(1),I_1,J_1,K_1,\e_1),(\lambda,I, \e)]  }
\nonumber \\
&& 
-\sum_{I_1}\sum_{J_1\neq I_1} S_{[(\lambda^I,(2),I_1,J_1,\e_1) (\lambda,I, \e)]  } + \sum_{I_1} j_{I_1}(j_{I_1}+1)(N-1)S_{\lambda, I, \e}\;\big).
\label{comm}
\ea
Our (more or less obvious) notation is as follows.
We have slightly abused notation and used the subscript $\lambda$ to denote the state $S_{\lambda, I,\e}$
with  intertwiner $C_{\lambda}$ despite the fact that the intertwiner change is due to the action of ${\hat q}^{-\frac{1}{2}}$ here as opposed to ${\hat q}^{-\frac{1}{3}}$ in (\ref{5.6}).
  As we have seen,  the only type of first transformation to contribute is   $(\lambda, I,\e)$ which sends $S$ to $S_{\lambda, I, \e}$. Recall that $S_{\lambda, I, \e}$ is obtained by a displacing the vertex structure
of $S$ at $v$ to the new vertex, here denoted by $v_{I,\e}$,  the intertwiner at $v_{I,\e}$ now being $C_{\lambda}$. This intertwiner changes due to the action of ${\hat q}^{-\frac{1}{2}}$ coming from the second constraint. 
Since the change in the intertwiner depends on the conically deformed vertex structure at $v_{I,\e}$, we denote  the resulting intertwiner by $C_{\lambda^I, \lambda}$. 
The notation in square brackets indicate that the transformation specified by the second round bracket is followed by that specified by the 
first bracket.
The second transformation can be any of:\\
\noindent (i) $(\lambda^I, I_1, \e_1)$ which displaces the vertex structure around $v_{I,\e}$ of $S_{\lambda, I, \e}$
along its $I_1$th edge by the amount $\e_1$ and changes the intertwiner $C_{\lambda}$ at this displaced vertex  to $C_{\lambda^I,\lambda}$, \\  
\noindent (ii) $(\lambda^I,(1),I_1,J_1,K_1,\e_1)$ which   changes the intertwiner $C_{\lambda}$ at $v_{I,\e}$ of $S_{\lambda, I, \e}$  to $C_{\lambda^I,\lambda}$ and then 
alters the vertex structure around $v_{I,\e}$ 
in accordance with Figure \ref{figmixed1}, or\\  
\noindent (iii) 
$(\lambda^I(2),I_1,J_1,\e_1)$ which changes the intertwiner $C_{\lambda}$ at $v_{I,\e}$ of $S_{\lambda, I, \e}$  to $C_{\lambda^I,\lambda}$ and then alters the vertex structure around $v_{I,\e}$  in accordance with Figure \ref{figmixed2}.

From the Appendix \ref{seca3} (see equation (\ref{a13})), the operator correspondent ${\hat O}(M, N)$ of the Poisson bracket between $H(M)$ and $H(N)$ can be regulated so as to act on $S$ as:
\ba
&&i{\hat O}_{\e}(M, N) S(A)=
(\frac{3}{4\pi})^2  \omega_0 
 \big(\sum_{I=1}^N\frac{ j_I(j_I+1)(S_{\eta,\lambda ,I, \e, \k} - S_{\eta,\lambda})}{\k}
\nonumber\\ 
&-&\sum_{I=1}^N\frac{  (\sum_{J\neq I}\sum_{K\neq I,J}S_{\eta,\lambda, (1)I,J,K,\e, \k }) +(\sum_{J\neq I} S_{\eta,\lambda, (2)I,J,\e, \k, \lambda_1, \lambda}) - j_I(j_I+1)(N-1)S_{\eta, \lambda }}{\k}\;\big)\;\;\;.\;\;\;\;\;\;\;
\label{rhsfinal}
\ea
Here $\omega_0$ is a quantity that depends solely on the lapses  and $\kappa$ is a constant which characterises the small loops $l_{IK,\e,\k}$ whose value we are free to choose (see the Appendix for details).
From the Appendix it turns out that each of the states, 
$S_{\eta,\lambda ,I, \e, \k}$,
$S_{\eta, \lambda,(1)I,J,K,\e, \k }$, $S_{\eta, \lambda, (2)I,J,\e, \k}$ are, respectively,  closely related to the states $S_{[(\lambda^I, I_1, \e_1),(\lambda,I, \e)]}$, $S_{ [(\lambda^I,(1),I_1,J_1,K_1,\e_1),(\lambda,I, \e)]  }$,
$S_{[(\lambda^I,(2),I_1,J_1,\e_1) (\lambda,I, \e)]  }$ in (\ref{comm}).
In particular if we repeat the consderations of this work with $\phi_{I,\e}$ chosen to be a {\em diffeomorphism}, each of the  former states turns out, respectively,  to be a diffeomorphic image of each of the latter.

\subsection{\label{sec5.3} An unconventional view of the quantum constraint algebra}

Note that the lapse prefactor $(N(v) M(v_{I,\e})- M(v)N(v_{I,\e})$ in (\ref{comm}) vanishes as $\e \rightarrow 0$. 
\footnote{This is expected on general grounds \cite{p3} for density weight 1 constraints and is the reason
that we have considered the higher density constraint in earlier sections. }
This vanishing is tied to the {\em continuity} of the lapse functions.  
Under the same assumption, it turns out that  $\omega_0$ also vanishes so that Assumption 1 of the Appendix \ref{seca3} is satisfied with $\omega_0=0$.

Let us now {\em assume} that the lapses can be discontinuous at {\em isolated} points. 
Let us also assume for the purposes of this section that {\em the transformations $\phi_{I,\e}$ are diffeomorphisms} instead of conical transformations.
Under these assumptions  and setting $\kappa^{-1} = \frac{\sum_{I}j_I(j_I+1)}{4}$ we find that 
that (a) the $\e \rightarrow 0$ limit of the  action of any diffeomorphism invariant distribution on the commutator (\ref{comm}) is the same as  its action on (\ref{rhsfinal})
and (b) this limit is non-trivial whenever the vertex $v$ coincides with an isolated point of discontinuity of at least one of the lapses.

This may possibly be viewed as an illustration of a non-trivial anomaly free action.
Since this is an unconventional point of view, we do not wish to unduly embellish it and restrict ourselves to the following remarks:\\

(1) 
Note that the commutator is non-trivial only if the displaced vertex  in the state $S_{\lambda, I,\e}$ is non-degenerate. The fact that the deformed vertex structure is diffeomorphic to the 
undeformed  one implies that the non-degeneracy of the former guarantees  that of the latter if the inverse deteriminant is defined in a diffeomorphism covariant way, as is usually done.
\\

(2) The equality can be viewed as the equality of continuum limits in a operator topology defined by a family of semi-norms, each semi-norm defined by a pair of elements, the first being a diffeomorphism 
invariant distribution and the second a spin network state of the type $S$ (see \cite{p1} for this viewpoint). Alternatively one could try to weaken the Uniform Rovelli Smolin Operator topology \cite{qsd,ttbook}
to allow a non-uniformity with respect to the lapse evaluation. It would be of interest to make the notion of  this `weakened' topology precise. We also feel that a habitat based demonstration 
should be possible and we leave this for future work.
\\

(3) If we implement $\phi_{I,\e}$ as a conical deformation, the non-degeneracy condition is preserved for the displaced vertex $S_{\lambda,I,\e}$ if we use the Rovelli-Smolin volume in conjunction with 
the Tikhonov regularization (see Footnote \ref{fntycho} and \cite{eugenio}). The  conical deformation is an example of an `extended diffeomorphism' \cite{rf}. Hence it seems likely that the non-trivial equality,\\ 
$\lim_{\e\rightarrow 0} \lim_{\e_1\rightarrow 0}\Phi ( [{\hat H}(M),{\hat H}(N)]_{\e_1, \e} S) = i\lim_{\e\rightarrow 0} \Phi ({\hat O}_{\e}(M, N) S)$, \\
holds for spin network states $S$ of the type considered in section \ref{sec5.1} and
distributions $\Phi$ which are invariant with respect to these singular diffeomorphisms if we use the Rovelli-Smolin volume operator. 
\\

(4) Note that the extended diffeomorphisms of \cite{rf} cannot be generated by smooth vector fields; some amount of non-smoothness is required of any putative generator.
From this point of view, the use of discontinuous lapse functions does not seem strange; if non-smooth shifts can be contemplated why not discontinuous lapses?
On the other hand, also note that the GR condition on $S$ is {\em not} preserved by extended diffeomorphisms.
\\

To summarise:  The use of unit density constraints avoids many of the technical complications which arise in the consideration of higher density constraints. 
However for smooth lapses the commutator between two Hamiltonian constraints as well as the operator correspondent of their Poisson bracket trivialise. Trivialization can be
avoided by the unconventional assumption of lapses with  discontinuities.  Then,
(i) if  these discontinuities are at isolated points,  
(ii) if we define a valid but unconventional regulation of the operator 
correspondent of the constraint Poisson bracket, and (iii) we implement the $\{\phi_{I,\e}\}$ as diffeomorphisms,\\
it is possible to define  a non-trivial continuum limit of this regularization so that it reproduces the $(-i)$ times the continuum limit of 
the commutator between the constraint operators.

\section{\label{sec6}Discussion}

In this work we have attempted to combine  the standard construction methods of LQG \cite{qsd,ttbook}  together with  geometric insights into  the classical dynamics of Euclidean General Relativity \cite{aame} in order to construct an action 
of the Hamiltonian constraint operator. The geometric interpretation of the classical dynamics derives from a rewriting of the Einstein equations in which time evolution of phase space fields
can be understood as Lie derivatives, suitably generalised, with respect to  a  Lie algebra valued, phase space dependent, spatial vector field called the Electric Shift \cite{aame} .
Our derivation accords a central role to the corresponding Electric Shift operator. 
%
Folding the action of the Electric Shift operator  into that of the constraint operator results in the action (\ref{3.7}) of section \ref{sec3}. The derivation of 
this action is our main 
result.\footnote{Note that while we derived (\ref{3.7}) under the assumption of lapse support around a single vertex, the derivation trivially
extends to the case of arbitrary lapse support. The constraint action is then a sum over contributions of the type (\ref{3.7}), one for every non-degenerate vertex of $S$. Similarly the results of sections \ref{sec4} and \ref{sec5}
also trivially extend to arbitrary lapses by summing over all vertex contributions.}
Its simplicity is tied to a crucial step in our derivation in which we combine   the action of the curvature term 
with that of the electric field operators in a delicate manner 
without recourse to the relatively brutal substitution  of curvature components by traces of holonomies with respect to $\tau_i$. 
In this regard our treatment of the crucial curvature term is closer to that of Thiemann's implementation of the
`right hand side' of the constraint commutator, ${\hat O}(M,N)$, in \cite{qsd3,ttbook} than to standard treatments of this term in the  constraint operator itself.

The combined action of the curvature and electric field dependent operators naturally brings  the deformations $\{\phi_{I,\e}\}$ to the fore, 
each $\phi_{I,\e}$ defining a deformation of the vertex structure at $v$ along the edge $e_I$ emanating therefrom.
However, the resulting action (\ref{3.7}) can also
be viewed in terms of a specific choice of holonomy approximants to the curvature of the connection.
In this specific choice, 
these approximants are labelled by spin quantum numbers 
which are tailored to the spin labels of the edges of the spin network state being acted upon.  The qualitative reason for this `dynamical' choice of spin labels is as follows. The  
 transformations $\phi_{I,\e}$ deform the edges at the vertex of interest without changing their spin labels. In order to implement such a deformation through the action of holonomy opertaors, it 
 would be necessary to eliminate the original edges by (inverse) holonomy multiplication and re-introduce their (deformed) images also by holonomy multiplication. Clearly this would need these
 regulating holonomies to be labelled by  the spin labels of the edges. Therefore, while in our derivation this feature can be traced to our specific visualization of $\phi_{I,\e}$, the feature
 itself (and the consequent elimination of Perez's spin ambiguity \cite{perez}) seems to be quite robust and valid for any reasonable implementation of $\phi_{I,\e}$.

In the standard construction of the Hamiltonian constraint \cite{qsd,ttbook} most of the remaining ambiguity is  tied to the detailed choice of  loops  (for e.g. their topology, placement and routing)
for the regulating holonomies. In our derivation  this is all encapsulated in our visualization of $\phi_{I,\e}$. 
Based on our intuition and experience with toy models, we have visualised each  $\phi_{I,\e}$ to act as a conical deformation so that its action falls into the class of extended diffeomorphisms
introduced by Fairbairn and Rovelli \cite{rf}. Another natural  visualizaton would be that of a diffeomorphism as in our brief speculative detour of sections \ref{sec5.2}  and \ref{sec5.3}.
It is remarkable that these natural possibilities give rise to the constraint action of section \ref{sec3},  which, purely in terms of edge placements, resembles the choices made in \cite{qsd,ttbook}.
In accordance with our view that these 
choices be confined through physical requirements, we further developed the constraint action of section \ref{sec3} into the  form
derived in section \ref{sec4} so as to confront the requirement of  a {\em non-trivial} anomaly free constraint algebra.  
\footnote{
A necessary condition for  non-triviality of the  commutator between a pair of Hamiltonian constraints is that the second constraint act on deformations generated by the first \cite{lm}. 
This ensures that the
second lapse is evaluated at a different vertex than the first and thereby yields  a result which is not symmetric under interchange of lapses thus preventing the antisymmetry of this combination
from trivialising the commutator. 
In contrast to the constraint action of section \ref{sec3}, the action of section \ref{sec4} constructs deformed and displaced vertex structures on which a second action
is not necessarily trivial.} 
This form closely resembles that in the $U(1)^3$ toy model \cite{p3}. However, preliminary calculations seem to indicate technical complications in any straightforward effort to generalise the 
considerations of \cite{p3} to the $SU(2)$ case.
Hence, we further developed the constraint action in section \ref{sec5}. The final action (\ref{5.6}) results in two distinct types of deformations.
The first  is  a conical deformation (and hence and extended diffeomorphism) and displacement of the original  vertex structure as a {\em whole} and the second deforms the vertex structure around the original vertex
`one edge at a time'. As a result, roughly speaking, the first piece is expected to a play a crucial role in any putative demonstration  of an anomaly free constraint algebra and the second piece is expected to 
seed propagation along the lines of \cite{ttme}.  Ongoing work consists of an investigation into these matters. 

We close with the following remarks which expand on our brief summary above and point to open issues:\\

\noindent (1) {\em Use of the GR property} (This remark may be skipped on a first reading as it concerns technicalities detailed in \cite{p3}): 
The GR property played an important role in the treatment of the constraint algebra in the $U(1)^3$ toy model context \cite{p3} and we anticipate that (some weakened version of) the GR property  may play an important role in 
analysing the constraint algebra with regard to the anomaly free property in the context of the $SU(2)$ case as well.
While we restricted attention to nondegenerate  vertices satisfying the GR property  in sections \ref{sec4} and \ref{sec5} 
there seems to be no obstruction to repeating the considerations of those sections  for general (nondegenerate) vertex structures.
The completion of such an exercise hinges on a specification of the transformation $\phi_{I,\e}$ on such vertices. In the GR case \cite{p3} it is straightforward to specify the deformation
in such a way that no unwanted intersections between the set of  deformed and undeformed edges are created. For a non-GR vertex with no  colinear edge tangents, all our considerations would go
through 
unchanged except that the specification of  the action of $\phi_{I,\e}$
on such vertices 
would have to confront issues similar to those by Thiemann in \cite{qsd} with regard to routing of the deformed edges so that no unwanted intersections arise.
This non-colinearity property is expected to be preserved by the mixed action of section \ref{sec5}.
Hence for the purposes of probing the constraint algebra one may restrict attention to such vertices and view, similar to the GR condition \cite{p3}, such a restriction 
as the analog of a nondegeneracy property in the quantum setting \cite{p3}. 
If sets of edges with colinear or anticolinear tangents are present new issues such as the increasing valence problem described in section \ref{sec4} may arise and need to be resolved, with a 
satisfactory generalisation of the mixed action offering a likely resolution.
\\


\noindent (2) {\em The choice of Volume operator}: The volume operator plays a key role in the construction of the 
inverse metric determinant operator. Preservation of vertex non-degeneracy with respect to the inverse metric determinant operator under the deformations $\{\phi_{I,\e}\}$ is likely to be a key ingredient in 
any proof of the anomaly free property (see the discussion of `eternal non-degeneracy' in \cite{p3}).
If these deformations were diffeomorphisms, the diffeomorphism covariance of any acceptable definition of the inverse metric determinant would ensure preservation of such non-degeneracy.
Since we have chosen the $\{\phi_{I,\e}\}$ to be conical deformations and since conical deformations fall into the class of extended diffeomorphisms defined by Fairbairn and Rovelli in \cite{rf}, preservation of non-degeneracy
under such deformations would require a construction of the inverse determinant metric which interacted well with such extended diffeomorphisms. For example, if one used the Rovelli-Smolin volume operator \cite{rsvol}
in conjunction with the Tikhonov regularization (see Footnote \ref{fntycho} and \cite{eugenio}), non-degeneracy would be preserved under conical deformations. As mentioned in section \ref{sec5.3}, the proposed use  of  lapses which are discontinuous at isolated points to probe 
the constraint algebra of unit density constraints seems to be in coherence with the Winston-Fairbairn proposal to  extend the group of diffeomorphisms to maps which are diffeomorphisms almost everywhere except at 
isolated points \cite{rf,carlopvt}.
\\

\noindent (3) {\em The properties of Propagation and Anomaly free action}: 
From \cite{ttme}, a key ingredient for a quantum dynamics  to display propagation is the existence of `children of non-unique parentage'.  Roughly speaking, by a child of non-unique parentage we mean a spin network which can be 
generated by the action of the constraint on two or more  diffeomorphically distinct `parent' spin networks i.e. a given deformed spin network does not  arise from the constraint action on a unique `parent' state.
It is then very likely that arguments similar to those in \cite{ttme} and applied to  children of type $S_{\lambda (2), I,J, K, \e}$  will establish vigorous propagation for the mixed action dynamics of section \ref{sec5}.
On the other hand, notwithstanding the discussion in section \ref{sec5.3}, a comprehensive and satisfactory demonstration of  a non-trivial anomaly free 
constraint algebra constitutes an open problem. In particular, relative to section \ref{sec5.3}, we would like to provide such a demonstration without recourse to discontinuous 
lapses and also without recourse to an unconventional regularization of  ${\hat O}(M,N)$. Thus we would like such a demonstration to be based on higher density constraints using and generalising the techniques developed in 
\cite{p3}, and, we would like  the operator correspondent of the constraint Poisson bracket,  ${\hat O}(M,N)$,  to act as a (perhaps, extended) diffeomorphism.
Investigations into this issue constitute work in progress.
\\

\noindent (4) {\em A comment on the $\alpha_\e$ expansion}:  In this work we have assumed that the small loop area parameter $\alpha_\e$ is of $O(\e^2)$. However, the action of $\phi_{I,\e}$
in \cite{p3} corresponds to small loops for which  $\alpha_\e$ is of a higher order than $\e^2$ and we plan to combine the results of this work together with the methods of \cite{p3}
in an effort to demonstrate the existence of an anomaly free, propagating constraint action for Euclidean LQG.
While a detailed discussion of this  and related tensions between the structures underlying \cite{p3} and the considerations in this work is beyond the scope of this paper, 
it is pertinent to  reiterate our general viewpoint: Our  viewpoint (motivated by the central role of the electric shift in the classical theory) is to treat 
all  heuristic
argumentation as a crutch whose purpose is   to obtain a geometrically compelling constraint action in which the deformations  $\phi_{I,\e}$ play a key role and 
whose justification would only be {\em a posteriori} through a putative demonstration of its  consistency with an anomaly free
constraint algebra and the property of propagation.
\\

\section*{Acknowledgements:} I am very grateful to Fernando Barbero for  his comments on a draft version of this manuscript, for his constant encouragement  and for his kind help with the figures. 
I thank Abhay Ashtekar, Alejandro Perez and Jorge Pullin for  discussions and comments, and 
their support and encouragement. I thank Rodolfo Gambini, Seth Major and members of the LSU Relativity Group, especially Andrea Dapor, for discussions and comments during the course of   online presentations
of this work. I thank Christian Fleischhack and Carlo Rovelli for their comments on extended diffeomorphisms.

\appendix

\section{\label{seca1}Explicit segregation of $O(\alpha_\e)$ terms in (\ref{4.8})}
Note that from Footnote \ref{fnorder}, the combination of the two terms in the curly brackets of (\ref{4.8}) is of order $\alpha_{\e}$.
We further manipulate (\ref{4.8}) so as to explicitly segregate the order $\alpha_{\e}$ contributions in  each of the two terms as follows.
To do so we note that:
\ba
&& C_{\lambda}^{D_1..  .D_N} (\prod_{J\neq I} (h_{e_J}(1,t_{\e,J,I}) h_{\phi_{I,\e}(e_J)}(t_{\e,J,I}, 0)
)^{B_J}{}_{D_J} )  h_{e_I}(1,\e_I)^{B_I}{}_{D_I}  \\
&=& C_{\lambda}^{D_1..  .D_N} (\prod_{J\neq I} (h_{e_J}(1,0) h_{l_{IJ,\e}} (h_{e_I}(\e_I,0))^{-1}
)^{B_J}{}_{D_J} )  h_{e_I}(1,\e_I)^{B_I}{}_{D_I} \\
&=&C_{\lambda}^{E_1...E_N} (\prod_{J\neq I} ( h_{e_J}(1,0)  h_{ l_{IJ,\e} } 
)^{B_J}{}_{E_J} ) h_{e_I}(\e_I,0)^{D_I}{}_{E_I} h_{e_I}(1,\e_I)^{B_I}{}_{D_I} \\
&=& C_{\lambda}^{E_1..  .E_N} (\prod_{J\neq I} (h_{e_J}(1,0) h_{l_{IJ,\e}}
)^{B_J}{}_{E_J} ) h_{e_I}(1,0)^{B_I}{}_{E_I}  \label{4.9a}\\
&=&  C_{\lambda}^{E_1..  .E_N} (\prod_{J=1}^N (h_{e_J}(1,0))^{B_J}{}_{E_J} ) + O(\alpha_e) 
\label{4.9}
\ea
where we have used (\ref{deflij}) in the second line, (\ref{cgginv2}) with $g\equiv (h_{e_I}(\e,0))^{-1}$ in the third,  and used the fact that $h_{l_{IJ,\e}} = {\bf 1} + O(\alpha_{\e})$ in the last line.
Using (\ref{deflij}), (\ref{cgginv2}) in an identical manner and again noting that $h_{l_{IJ,\e}} = {\bf 1} + O(\alpha_{\e})$, we have that:
\ba
&& C_{\lambda}^{D_1..D_N} (\prod_{K\neq I,J } h_{e_K}(1,t_{\e,K,I}) h_{\phi_{I,\e}(e_K)}(t_{\e,K,I}, 0)
)^{B_K}{}_{D_K} )      \nonumber\\
&&(h_{e_J}(1,0)\tau_j 
(h_{e_J}(t_{\e,J,I}, 0)   )^{-1} h_{\phi_{I,\e}(e_J)}(t_{\e,J,I}, 0)
)^{B_J}{}_{D_J} 
(h_{e_I}(1,0)\tau_j)^{B_I}{}_{A_I}
((h_{e_I}(\e_I,0))^{-1})^{A_I}{}_{D_I}\nonumber \\
=&&C_{\lambda}^{D_1..D_N} (\prod_{K\neq I,J } h_{e_K}(1,0) h_{l_{IK,\e}} (h_{e_I}(\e,0))^{-1}
)^{B_K}{}_{D_K} )      \nonumber\\
&&(h_{e_J}(1,0)\tau_j 
h_{l_{IJ,\e}} (h_{e_I}(\e,0))^{-1}
)^{B_J}{}_{D_J} 
(h_{e_I}(1,0)\tau_j)^{B_I}{}_{A_I}
((h_{e_I}(\e_I,0))^{-1})^{A_I}{}_{D_I} \\
=&&C_{\lambda}^{D_1..D_N} (\prod_{K\neq I,J } h_{e_K}(1,0) h_{l_{IK,\e}} 
)^{B_K}{}_{D_K} )      
(h_{e_J}(1,0)\tau_j 
h_{l_{IJ,\e}} 
)^{B_J}{}_{D_J} 
(h_{e_I}(1,0)\tau_j)^{B_I}{}_{D_I}
\label{4.10a}
\\
=&&C_{\lambda}^{D_1..D_N} (\prod_{K\neq I,J } h_{e_K}(1,0) 
)^{B_K}{}_{D_K} )      
(h_{e_J}(1,0)\tau_j  
)^{B_J}{}_{D_J} 
(h_{e_I}(1,0)\tau_j)^{B_I}{}_{D_I} + O(\alpha_\e)\nonumber\\
\label{4.10}
\ea
Equation (\ref{infgginv}) implies that 
\ba
&&\sum_{J\neq I} C_{\lambda}^{D_1..D_N} (\prod_{K\neq I,J } h_{e_K}(1,0) 
)^{B_K}{}_{D_K} )      
(h_{e_J}(1,0)\tau_j  
)^{B_J}{}_{D_J} 
(h_{e_I}(1,0)\tau_j)^{B_I}{}_{D_I} \nonumber\\
&=&
-C_{\lambda}^{D_1..D_{I-1}C_ID_{I+1}..D_N} (\prod_{J\neq I } h_{e_J}(1,0) 
)^{B_J}{}_{D_J} )      
 (\tau_j)^{D_I}{}_{C_I}
(h_{e_I}(1,0)\tau_j)^{B_I}{}_{D_I} \nonumber \\
&=&
j_I(j_I+1) C_{\lambda}^{D_1...D_N} (\prod_{J} h_{e_J}(1,0) 
)^{B_J}{}_{D_J} )      
\label{4.11}
\ea
Equations (\ref{4.10}) and (\ref{4.11}) imply that:
\ba
&& \sum_{J\neq I} C_{\lambda}^{D_1..D_N} (\prod_{K\neq I,J } h_{e_K}(1,t_{\e,K,I}) h_{\phi_{I,\e}(e_K)}(t_{\e,K,I}, 0)
)^{B_K}{}_{D_K} )      \nonumber\\
&&(h_{e_J}(1,0)\tau_j 
(h_{e_J}(t_{\e,J,I}, 0)   )^{-1} h_{\phi_{I,\e}(e_J)}(t_{\e,J,I}, 0)
)^{B_J}{}_{D_J} 
(h_{e_I}(1,0)\tau_j)^{B_I}{}_{A_I}
((h_{e_I}(\e_I,0))^{-1})^{A_I}{}_{D_I}\nonumber \\
=&& j_I(j_I+1) C_{\lambda}^{D_1...D_N} (\prod_{J} h_{e_J}(1,0) 
)^{B_J}{}_{D_J} )  +O(\alpha_{\e})
\label{4.12}
\ea
Equations (\ref{4.9}) and (\ref{4.12}) imply that the right hand side of (\ref{4.12}) corresponds to the $O(1)$ part of each of the curly brackets.
Hence we may subtract out the $O(1)$ contributions to each of these lines and write (\ref{4.8})  in the form:
\ba
&&{\hat H}_{\epsilon}(N) S(A) := \frac{3}{8\pi}N(x(v)) \frac{1}{\e}\sum_{I=1}^N    S_{rest}{}_{B_I..B_N}(A) 
\nonumber\\
& \big(& \{j_I(j_I+1) C_{\lambda}^{D_1..  .D_N} (\prod_{J\neq I} (h_{e_J}(1,t_{\e,J,I} h_{\phi_{I,\e}(e_J)}(t_{\e,J,I}, 0)
)^{B_J}{}_{D_J} )  h_{e_I}(1,\e_I)^{B_I}{}_{D_I} 
\nonumber \\
&&
- j_I(j_I+1)C_{\lambda}^{D_1..  .D_N} (\prod_{J=1}^N h_{e_J}(1,0)^{B_J}{}_{D_J}) \} 
\nonumber 
\\
&-&\{ \sum_{J\neq I} C_{\lambda}^{D_1..D_N} (\prod_{K\neq I,J }( h_{e_K}(1,t_{\e,K,I}) h_{\phi_{I,\e}(e_K)}(t_{\e,K,I}, 0)
)^{B_K}{}_{D_K} )      \nonumber\\
&&(h_{e_J}(1,0)\tau_j 
(h_{e_J}(t_{\e,J,I}, 0)   )^{-1} h_{\phi_{I,\e}(e_J)}(t_{\e,J,I}, 0)
)^{B_J}{}_{D_J} 
(h_{e_I}(1,0)\tau_j)^{B_I}{}_{A_I}
((h_{e_I}(\e_I,0))^{-1})^{A_I}{}_{D_I} 
\nonumber
\\
&&
- j_I(j_I+1) C_{\lambda}^{D_1..  .D_N}( \prod_{J=1}^N h_{e_J}(1,0)^{B_J}{}_{D_J}) \}\;\;\;\big) 
\label{4.13}
\ea
where the terms in curly brackets are now each of $O(\alpha_\e)$. Note that the subtraction term in each curly bracket, when combined with $S_{rest}{}_{B_I..B_N}(A)$ yields exactly $S_{\lambda}(A)$. Hence we may write
(\ref{4.13}) as:
\be 
{\hat H}_{\epsilon}(N) S(A) := \frac{3}{8\pi}N(x(v)) \big(\sum_{I=1}^N\frac{\{ j_I(j_I+1)(S_{\lambda,I, \e} - S_{\lambda})\}}{\e} + \sum_{I=1}^N(\frac{\{\sum_{J\neq I} S_{\lambda,I,J,\e}  - j_I(j_I+1)S_{\lambda}\}}{\e}\big)
\label{4.14a}
\ee
where we reproduce the definitions (\ref{sie}), (\ref{sije}) of section \ref{sec4}) as (\ref{siea}),(\ref{sijea}) below  for the convenience of the reader:
\be
S_{\lambda,I, \e} := C_{\lambda}^{D_1..  .D_N} (\prod_{J\neq I} (h_{e_J}(1,t_{\e,J,I}) h_{\phi_{I,\e}(e_J)}(t_{\e,J,I}, 0)
)^{B_J}{}_{D_J} )  h_{e_I}(1,\e_I)^{B_I}{}_{D_I} S_{rest\;B_1..B_N},
\label{siea}
\ee
\ba
&&S_{\lambda,I,J,\e} := 
C_{\lambda}^{D_1..D_N} (\prod_{K\neq I,J } (h_{e_K}(1,t_{\e,K,I}) h_{\phi_{I,\e}(e_K)}(t_{\e,K,I}, 0)
)^{B_K}{}_{D_K} )      \nonumber\\
&&(h_{e_J}(1,0)\tau_j 
(h_{e_J}(t_{\e,J,I}, 0)   )^{-1} h_{\phi_{I,\e}(e_J)}(t_{\e,J,I}, 0)
)^{B_J}{}_{D_J} 
(h_{e_I}(1,0)\tau_j)^{B_I}{}_{A_I}
((h_{e_I}(\e_I,0))^{-1})^{A_I}{}_{D_I} S_{rest\;B_1..B_N}\;\;\;\;\;\;\;\;\;\;\;
\label{sijea}
\ea

\section{\label{seca2.1} A useful Lemma}
{\bf Lemma}: Consider the matrix expression $ f =( h^{(j)}_{e}(A) )^{-1} \tau_i h^{(j)}_e(A)$ where $h_e(A)$ is an edge holonomy of the connection $A$ along the edge $e$ and $h^{(j)}_e(A), \tau_i$ are in a spin $j$ representation.
Then the Clebsch-Gordon decomposition of $f$ has contributions only from spin 1.
\\

\noindent{\em Proof}: We have that $h^{(j)}_e (A) =:g^{(j)}$  where $g^{(j)}$ is the spin $j$ matrix representative of the element $g\in SU(2)$.
Since $\tau_i$ is the  spin $j$ representative of the $j$th generator of $su(2)$ its index $j$ behaves as vector index under the action of $SU(2)$ transformations so that
\be
( g^{(j)})^{-1} \tau_i g^{(j)} =  R^k{}_i (g) \tau_k .
\ee
where the rotation matrix  $R^k{}_i(g)$ bears the interpretation of the spin 1 representative of $g$.  
It follows that $R(g) = h^{(1)}_e(A)$  where we have suppressed the matrix indices and where the superscript $(1)$ indicates that the holonomy is in the spin 1 representation.
This completes the proof.

\section{\label{seca2}A more general form of Hamiltonian constraint action through conical type deformations}

In this section we  change each of the two terms 
in (\ref{4.14}) through the addition of higher order terms in $\alpha_\e$ and thereby obtain a general form of the Hamiltonian constraint which encompasses (\ref{4.8}), (\ref{4.14}) and which may, conceivably, 
be of use in  future attempts at showing consistency with an anomaly free constraint algebra.

We define the loop $l^{c_I}_{IJ,\e}$ as follows:\\
(a) Similar to $l_{IJ,\e}$, the loop   $l^{c_I}_{IJ,\e}$ is  a `triangle' with two sides along $e_I$ and $e_J$ emanating from its vertex $v$ and a third side connecting the vertices on $e_I,e_J$
with each other.\\
(b) We adjust the placement of the vertices on the sides $e_J,e_I$  by a coordinate distance of  $O(\e)$ so that the area of the loop $l^{c_I}_{IJ,\e}$ is now $c_I \alpha_{IJ,\e}$.\\
It follows from standard holonomy expansion that:
\be
h_{ l^{c_I}_{IJ,\e}}(A) -{\bf 1} = -  ({\hat e}^a_I {\hat e}^b_J F_{ab}^i (v)\tau_i) c_I \alpha_{IJ,\e}  + O(\alpha_{IJ,\e}^2) = c_I (h_{ l_{IJ,\e}}(A) -{\bf 1})+ O(\alpha_{IJ,\e}^2) 
\label{hlcije}
\ee
Equations  (\ref{hlcije}),  and (\ref{4.9a}), (\ref{4.9}) imply that:
\ba
&& a_I\left(C_{\lambda}^{E_1..  .E_N} (\prod_{J\neq I} (h_{e_J}(1,0) h_{l_{IJ,\e}}
)^{B_J}{}_{E_J} ) h_{e_I}(1,0)^{B_I}{}_{E_I} 
-  C_{\lambda}^{E_1..  .E_N} (\prod_{J=1}^N (h_{e_J}(1,0))^{B_J}{}_{E_J} )\right)
\nonumber\\
&&
= C_{\lambda}^{E_1..  .E_N} (\prod_{J\neq I} (h_{e_J}(1,0) h_{l^{a_I}_{IJ,\e}}
)^{B_J}{}_{E_J} ) h_{e_I}(1,0)^{B_I}{}_{E_I} 
-  C_{\lambda}^{E_1..  .E_N} (\prod_{J=1}^N (h_{e_J}(1,0))^{B_J}{}_{E_J}  + O(\alpha_{\e}^2)\;\;\;\;\;\;\;\;\;\;
\label{4.9ci}
\ea
Similarly, equations (\ref{hlcije}) and (\ref{4.10a}), (\ref{4.10}) imply that:
\ba
&& b_I\big(C_{\lambda}^{D_1..D_N} (\prod_{K\neq I,J } h_{e_K}(1,0) h_{l_{IK,\e}} 
)^{B_K}{}_{D_K} )      
(h_{e_J}(1,0)\tau_j 
h_{l_{IJ,\e}} 
)^{B_J}{}_{D_J} 
(h_{e_I}(1,0)\tau_j)^{B_I}{}_{D_I}
\nonumber\\
&-& C_{\lambda}^{D_1..D_N} (\prod_{K\neq I,J } h_{e_K}(1,0) 
)^{B_K}{}_{D_K} )      
(h_{e_J}(1,0)\tau_j  
)^{B_J}{}_{D_J} 
(h_{e_I}(1,0)\tau_j)^{B_I}{}_{D_I}\big) \nonumber\\
&=&
C_{\lambda}^{D_1..D_N} (\prod_{K\neq I,J } h_{e_K}(1,0) h_{l^{b_I}_{I,K,\e}} 
)^{B_K}{}_{D_K} )      
(h_{e_J}(1,0)\tau_j 
h_{l^{b_I}_{IJ,\e}} 
)^{B_J}{}_{D_J} 
(h_{e_I}(1,0)\tau_j)^{B_I}{}_{D_I}
\nonumber\\
&-& C_{\lambda}^{D_1..D_N} (\prod_{K\neq I,J } h_{e_K}(1,0) 
)^{B_K}{}_{D_K} )      
(h_{e_J}(1,0)\tau_j  
)^{B_J}{}_{D_J} 
(h_{e_I}(1,0)\tau_j)^{B_I}{}_{D_I} + O(\alpha_{\e}^2)
\label{4.10ci}
\ea
Next note that the right hand side of (\ref{sie}) (modulo $S_{rest}$)  can be replaced by (\ref{4.9a}). Replacing the loops $l_{IJ,\e}$ in the resulting expression
for (\ref{sie}) by  the loops $l^{a_I}_{IJ,\e}$, we obtain an expression which we denote by $S_{\lambda, I, a_I, \e}$. 
Similarly the right hand side of (\ref{sije})  can be replaced by (\ref{4.10a}). Replacing the loops $l_{IJ,\e}$ in the resulting expression
for (\ref{sije}) by  the loops $l^{b_I}_{IJ,\e}$, we obtain an expression which we denote by $S_{\lambda, I, J, b_I, \e}$. Equations (\ref{4.9ci}) and (\ref{4.10ci}) 
then imply the action 
\be 
{\hat H}_{\epsilon}(N) S(A) := \frac{3}{8\pi}N(x(v) \left(\sum_{I=1}^N\frac{ j_I(j_I+1)(S_{\lambda,I, \e} - S_{\lambda})}{a_I\e} + \sum_{I=1}^N\frac{(\sum_{J\neq I} S_{\lambda,I,J,\e})  - j_I(j_I+1)S_{\lambda}}{b_I\e}\;\right)
\label{4.15}
\ee
agrees with the (\ref{4.14}) to leading order i.e. upto terms of $O(\alpha_\e^2)$. Hence (\ref{4.15}) constitutes a valid approximant to the Hamiltonian constraint.
Equation (\ref{4.15}) is our final result.

\section{\label{seca3} The operator correspondent of $\{H(M), H(N)\}$}
On the Gauss Law constraint surface (\ref{classhh}) for density 1 constraints we have that:
\be
\{H[M],H[N]\} = \int d^3x \omega_a q^{-1}E^{ia}E_{i}^{b} E^c_kF_{bc}^k := O(M,N)
\label{a1}
\ee
where 
\be
\omega_a:= N\partial_aM-M\partial_aN
\label{a2}
\ee
In this section we  construct the action of a regulated operator correspondent of (\ref{a1}) on the gauge invariant spin network $S(A)$ at its non-degenerate vertex $v$ with $v$ assumed to be GR.  We shall use the notation of the main text, 
assume  that the supports of the lapses $N,M$
include only a single nondegenerate vertex $v$ of $S$ and 
choose an operator ordering in which the regulated inverse metric determinant acts first. Similar to the case of the regulated Hamiltonian constraint, the only nontrivial contribution to the regulated  operator action then comes from 
a ball $B_\e(v)$ of coordinate size $\e$ around $v$. Accordingly we define:
\be
{\hat O}_\e(M,N) S(A): =  \frac{4\pi}{3}\e^3 (\omega_{a}(v))_\e ((F_{bc}^k)(v))_\e({\hat E}^{ia})_\e ({\hat E}^c_{k})_\e ({\hat E}_{i}^{b})_\e( {\hat q}^{-1/2})_\e ( {\hat q}^{-1/2})_\e S(A)
\ee
where we have ommitted the `hat' on the regulated curvature term as it acts by multiplication.
The regulated inverse metric determinant and electric field operators above are defined as follows.
${\hat q}^{-1/2}$ is obtained from the Volume operator and we expect   $\e^{-3} {\hat q}^{-1/2}_\e$ to be a well defined operator which only changes the intertwiner. We denote:
\ba
\e^{-3} {\hat q}^{-1/2}_\e S =: S_{\lambda}, \label{a4} \\
\e^{-3} {\hat q}^{-1/2}_\e S_{\lambda}  =: S_{\eta, \lambda}
\label{a5}
\ea
where  $S_{\eta, \lambda}$ differs from $S_{\lambda}$  by a change in the intertwiner $C_{\lambda}$ of the former to the new intertwiner $C_{\eta,\lambda}$.
Each of the electric field operators is regulated in a manner similar to the main text 
by averaging its action over the ball $B_\e$.  Thus, if ${\bar S}$ is a gauge invariant spin network and $B_\e$ contains only a single  nondegenerate vertex $v$ of ${\bar S}$, it is straightforward to repeat 
a derivation similar to that of (\ref{qshift}) and obtain:
\ba
({\hat E}^f_{m})_\e {\bar S} &:=& \frac{3}{4\pi\e^3} \int_{B_\e(v)} d^3x {\hat E}^f_{m}(x) {\bar S}\label{a6}\\
&=& i\frac{3}{4\pi\e^2} \sum_{I} {\hat e}_I^f {\hat X}_{I,m} {\bar S}
\ea
where ${\hat e}_I^f$ is the unit coordinate tangent to the $I$th edge at the vertex $v$.
It follows that:
\ba
i{\hat O}_\e(M,N) S(A) &: =&  (\frac{3}{4\pi})^2 \sum_{I,J,K} (\e{\hat e}_J^a\omega_{a}) (\e^2F_{bc}^k)(v) {\hat e}^c_K {\hat e}^b_I {\hat X}^i_{J} {\hat X}_{k,K}  {\hat X}_{i,I}S_{\eta, \lambda}
\label{a7.0} \\
&=& -(\frac{3}{4\pi})^2 \left(  \sum_{I} \sum_{K\neq I} O_{IIK} +  \sum_{I} \sum_{J\neq I}O_{IJJ} +  \sum_{I} \sum_{J\neq I} \sum_{K\neq I,J} O_{IJK}\right)
\label{a7.1}
\ea
where 
\be
O_{IJK}:=(\e{\hat e}_J^a\omega_{a}) (-\e^2F_{bc}^k)(v) {\hat e}^c_K {\hat e}^b_I {\hat X}^i_{J} {\hat X}_{k,K}  {\hat X}_{i,I}S_{\eta, \lambda}.
\label{a8}
\ee
and we have used the antisymmetry of $F_{bc}^k$ to drop the $K=I$ terms in the sum in (\ref{a7.0}).
Next we define  regulated versions of $(\e{\hat e}_J^a\omega_{a})$, $(\e^2F_{bc}^k)(v) {\hat e}^c_K {\hat e}^b_I$ through:
\be
(\e{\hat e}_J^a\omega_{a})_\e := N(v)M(v^a+ \e{\hat e}^a_J) - M(v)N(v^a+ \e{\hat e}^a_J), 
\label{a9}
\ee
where $v^a+ \e{\hat e}^a_J$ denotes the coordinates of the point obtained by traversing a coordinate distance $\e$ along the direction ${\hat e}^a_J$ from $v$, and,
\be
(-\e^2F_{bc}^k)(v) {\hat e}^c_K {\hat e}^b_I\tau_k := \frac{ h_{l_{IK,\e,\kappa}} -{\bf 1}}{\kappa} + O(\alpha_\e^2)
\label{flikc}
\ee
where $l_{IK,\e \kappa}$ is a triangular loop similar to $l_{IK,\e}$ in the main text except that it has area $\kappa\e^2$ for some constant $\kappa$ to be specified later.
Thus $l_{IK,\e, \kappa}$ traverses $e_I$ from $v$ at $t_I=0$ to some $v_{I,\e,\k}$ which lies at parameter value $t_I= \e_{I,\k}$ , then along an edge  from $v_{I,\e,\k}$ to a point on the  edge $e_K$ at some parameter value
$t_{\e,K,I,\kappa}$ and then from this point back to $v$ along $e_K$ in the inward direction. In addition 
we assume
\footnote{While this assumption seems reasonable to us, an explicit construction of such a diffeomorphism is desirable. We leave this for future work.}
that  the loops  $\{l_{IK,\kappa\e}, \}$ can be chosen in accordance with { Choice 1} below:\\

\noindent{\bf Choice 1}:  
The loops $\{l_{IK,\e, \k},\;K\neq I\}$ are chosen to be diffeomorphic images of their counterparts $\{l_{IK,\e},\; K\neq I\}$ by some diffeomorphism $\psi_{I,\e,\kappa}$
such that 
\be
l_{IK,\e,\k} = (e_K(t_{\e,K,I,\k}, 0)^{-1} \circ \phi_{I,\e,\k}(e_K)(t_{\e,K,I,\k}, 0) \circ e_{I}(\e_{I,\k}, 0)\label{likk}\\
\ee
where we have defined
\be
\phi_{I, \e, \kappa}  := \psi_{I, \e, \kappa} \circ \phi_{I, \e},
\label{defphik}
\ee
and where we have used the natural parameterization induced on the edge $\phi_{I,\e,\k}(e_K)(t_{\e,K,I,\k}, 0)$ by $\phi_{I, \e, \kappa}$ from the edge $e_K$.
Equation (\ref{likk}) implies that:
\be
h_{l_{IK,\e,\k}} = (h_{e_K}  (t_{\e,K,I,\k}, 0))^{-1}      
 h_{\phi_{I,\e,\k}(e_K)}(t_{\e,K,I,\k}, 0)h_{e_{I}}(\e_{I,\k}, 0) \label{hlikk}
\ee
We evaluate each of the terms in (\ref{a7.1}) in turn using (\ref{a9}), (\ref{flikc}),  (\ref{defphik}) and (\ref{hlikk}). 
When using (\ref{flikc}) we shall drop the $O(\alpha_\e^2)$ term.
We shall see that the first term can be modified by higher curvature terms in a manner similar to the `sum to product' transformation
of section \ref{sec3} so as to obtain a contribution similar  to that of the first set of terms in the constraint commutator, namely the 
$S_{[(\lambda^I, I_1, \e_1),(\lambda,I, \e)]}$ terms
in the second line of (\ref{comm}).
The remaining two sets of terms in  (\ref{a8}) will be seen to be analogs of  the 
$S_{ [(\lambda^I,(1),I_1,J_1,K_1,\e_1),(\lambda,I, \e)]  }, 
S_{[(\lambda^I,(2),I_1,J_1,\e_1) (\lambda,I, \e)]  }$
in the constraint commutator (\ref{comm}).
We have that:
\ba
O_{IIK}&:=&(\e{\hat e}_I^a\omega_{a})_\e (-\e^2F_{bc}^k)(v) {\hat e}^c_K {\hat e}^b_I {\hat X}_{i,I} {\hat X}_{k,K}  {\hat X}_{i,I}S_{\eta, \lambda} \label{a12.1}\\
&=& 
(\e{\hat e}_I^a\omega_{a})_\e (-\e^2F_{bc}^k)(v) {\hat e}^c_K {\hat e}^b_I {\hat X}_{k,K}  {\hat X}_{i,I}{\hat X}_{i,I}S_{\eta, \lambda} \label{a12.2}  \\
&=& -(j_I)(j_I+1)
(\e{\hat e}_I^a\omega_{a})_\e (-\e^2F_{bc}^k)(v) {\hat e}^c_K {\hat e}^b_I  {\hat X}_{k,K}  S_{\eta, \lambda}\label{a12.3}\\
&=& -(j_I)(j_I+1)(\e{\hat e}_I^a\omega_{a})_\e  (-\e^2F_{bc}^k)(v) {\hat e}^c_K {\hat e}^b_I
(h_{e_K}(1,0)\tau_k )^{\;B_K}_{\;\;A_K}   \frac{\partial S_{\eta, \lambda} \;\;\;\;}{\partial h_{e_K}(1,0)^{\;B_K}_{\;\;A_K}}
 \label{a12.41}\\
&=&
-(j_I)(j_I+1)(\e{\hat e}_I^a\omega_{a})_\e  
(h_{e_K}(1,0) (\frac{ h_{l_{IK,\e,\k}} -{\bf 1}}{\kappa})      )^{\;B_K}_{\;\;A_K}   \frac{\partial S_{\eta, \lambda} \;\;\;\;}{\partial h_{e_K}(1,0)^{\;B_K}_{\;\;A_K}} \label{a12.4}.\\
\ea
Above, we have used $K\neq I$ to move ${\hat X}_{iI}$ past ${\hat X}_{kK}$ in (\ref{a12.2})  and used (\ref{flikc}) in (\ref{a12.4}).
Next we sum over  $K$, and add higher order terms in small loop area to move from a sum to a product {\em exactly} as in section \ref{sec3.3}. It is then straightforward  to obtain:
\be
\sum_{K\neq I} O_{IIK} = -\kappa^{-1}(j_I)(j_I+1)(\e{\hat e}_I^a\omega_{a})_\e \{S_{\eta,\lambda, I, \e, \k} - S_{\eta,\lambda}\} + O(\e^2)
\ee
where 
\be
S_{\eta, \lambda,I, \e, \k} := C_{\eta, \lambda}^{D_1..  .D_N} (\prod_{K\neq I} (h_{e_K}(1,t_{\e,K,I,\k}) h_{\phi_{I,\e, \k}(e_K)}(t_{\e,K,I,\k}, 0)
)^{B_K}{}_{D_K} )  h_{e_I}(1,\e_{I,\k})^{B_I}{}_{D_I} S_{rest\;B_1..B_N},
\label{siek}
\ee
Note that $S_{\eta, \lambda,I, \e, \k}$ is given exactly by substituting $C_{\lambda}$ and  $\phi_{I,\e}$    
by $C_{\eta,\lambda}$ and  $\phi_{I,\e,\k}$ in     (\ref{sie}).

Next consider 
\ba
O_{IJJ}&:=&(\e{\hat e}_J^a\omega_{a})_\e (-\e^2F_{bc}^k)(v) {\hat e}^c_J {\hat e}^b_I {\hat X}^i_{J} {\hat X}_{k,J}  {\hat X}_{i,I}S_{\eta, \lambda} \label{a12.5}\\
&=&(\e{\hat e}_J^a\omega_{a})_\e  (-\e^2F_{bc}^k)(v) {\hat e}^c_J {\hat e}^b_I
(h_{e_J}(1,0)\tau^i\tau_k )^{\;B_J}_{\;\;A_J}   \frac{\partial {\hat X}_{i,I}S_{\eta,\lambda} \;\;\;\;}{\partial h_{e_J}(1,0)^{\;B_J}_{\;\;A_J}}\\
&=&  \k^{-1}
(\e{\hat e}_J^a\omega_{a})_\e  
(h_{e_J}(1,0)\tau^i (h_{l_{IJ,\e,\k}}-{\bf 1}) )^{\;B_J}_{\;\;A_J}   \frac{\partial {\hat X}_{i,I}S_{\eta,\lambda} \;\;\;\;}{\partial h_{e_J}(1,0)^{\;B_J}_{\;\;A_J}}
\label{a12.6}
\ea
Next, we sum (\ref{a12.6}) over $J$. Let us assume that \\

\noindent {\bf Assumption 1}: The $O(1)$ contribution to $(\e{\hat e}_J^a\omega_{a})_\e$ is {\em independent} of $J$ for small enough $\e$ i.e.
\be
(\e{\hat e}_J^a\omega_{a})_\e = \omega_0 + O(\e) 
\ee
where $\omega_0$ is independent of $\e$.
\\

It is then straightforward to see, using (\ref{infgginv}),   that the  sum over  ${\bf 1}$ contributions yields, to leading order in $\e$, the result  $(\e{\hat e}_J^a\omega_{a})_\e (j_I)(j_I+1) S_{\eta,\lambda}$
It follows that the final result can be written  to leading order in $\e$ as:
\be
O_{IJJ}=  \k^{-1} \omega_0 \sum_{J\neq I}
(S_{\eta,\lambda, (2),I, J, \e,\k} - (j_I)(j_I+1) S_{\eta, \lambda})
\label{s2ijek}
\ee
where 
$S_{\eta,\lambda, (2),I, J, \e,\k}$ is given exactly by substituting, in (\ref{s2}),  $C_{\lambda}$ and  $l_{IJ,\e}$    by
 $C_{\eta,\lambda}$ and  $l_{IJ,\e,\k}$  respectively.

Finally, proceeding in a similar manner, we have that:
\ba
O_{IJK}|_{I\neq J\neq K} &=& (\e{\hat e}_J^a\omega_{a})_\e (-\e^2F_{bc}^k)(v) {\hat e}^c_K {\hat e}^b_I {\hat X}^i_{J} {\hat X}_{k,K}  {\hat X}_{i,I}S_{\eta, \lambda} \\
&=&(\e{\hat e}_J^a\omega_{a})_\e  (-\e^2F_{bc}^k)(v) {\hat e}^c_K {\hat e}^b_I
(h_{e_K}(1,0)\tau_k )^{\;B_K}_{\;\;A_K}   \frac{\partial {\hat X}^i_{J} {\hat X}_{i,I}S_{\eta \lambda} \;\;\;\;}{\partial h_{e_J}(1,0)^{\;B_K}_{\;\;A_K}} \\
&=& \k^{-1}(\e{\hat e}_J^a\omega_{a})_\e 
(h_{e_K}(1,0)(h_{l_{IK,\e,\k}} -{\bf 1}) )^{\;B_K}_{\;\;A_K}   \frac{\partial {\hat X}^i_{J}{\hat X}_{i,I}S_{\eta, \lambda} \;\;\;\;}{\partial h_{e_J}(1,0)^{\;B_K}_{\;\;A_K}} 
\label{a12.7}
\ea
Summing the ${\bf 1}$ term over $K$ yields  $N-2$ identical terms. Under Assumption 1 above, we may then sum over $J$ and use (\ref{infgginv}).  The result is
$(N-2) (j_I)(j_I+1) S_{\eta, \lambda}$. It follows that the final result can be written to leading order in  $\e$ as:
\be
\sum_{J, K\neq J} O_{IJK}=  \k^{-1} \omega_0 (\sum_{J, K\neq J }
S_{\eta,\lambda, (1),I, J, K \e,\k}) - (N-2) (j_I)(j_I+1) S_{\eta, \lambda})
\label{s2ijkek}
\ee
where $S_{\eta,\lambda (1),I, J, K \e,\k}$  is given exactly by substituting,  in     (\ref{s1}), $S_{\lambda}$ and $l_{IK,\e}$   
by $S_{\eta,\lambda}$ and  $l_{IK,\e,\k}$ respectively.

Putting all this together we obtain our final result that to leading order in $\e$ we have that:
\ba
&&i{\hat O}_\e(N,M) S(A) =(\frac{3}{4\pi})^2  \omega_0 
 \big(\;\;\sum_{I=1}^N\frac{ j_I(j_I+1)(S_{\eta,\lambda ,I, \e, \k} - S_{\eta,\lambda})}{\k}
\nonumber\\ 
&-&\sum_{I=1}^N\frac{ (\sum_{J\neq I}\sum_{K\neq I,J}S_{\eta, \lambda,  (1)I,J,K,\e, \k }) +(\sum_{J\neq I} S_{ \eta, \lambda, (2)I,J,\e, \k}) - j_I(j_I+1)(N-1)S_{\eta, \lambda }}{\k}
\;\;\;\big)\;\;\;.\;\;\;\;\;\;\;
\label{a13}
\ea

\end{document}